\documentclass[aps,twocolumn,preprintnumbers,amsmath,amssymb]{revtex4-2}

\usepackage{graphicx}
\usepackage{dcolumn}
\usepackage{bm}
\usepackage{multirow}
\usepackage{color}
\usepackage{ulem}
\usepackage{amsmath}
\usepackage{gensymb}

\begin{document}
\title{Imaging Seebeck drift of excitons and trions in MoSe$_2$ monolayers}

\email{fabian.cadiz@polytechnique.edu}
\author{S. Park $^{1}$}
\author{B. Han $^{2}$}
\author{C. Boule $^{1}$}
\author{D. Paget $^{1}$}
\author{A. C. H.  Rowe $^{1}$}
\author{F. Sirotti $^{1}$}
\author{T. Taniguchi$^3$}
\author{K. Watanabe$^4$}
\author{C. Robert $^{2}$}
\author{L. Lombez $^{2}$}
\author{B. Urbaszek $^{2}$}
\author{X. Marie $^{2}$}
\author{F. Cadiz$^{1}$}

\affiliation{$^1$ Physique de la mati\` ere condens\' ee, Ecole Polytechnique, CNRS, IP Paris, 91128 Palaiseau, France}

\affiliation{$^2$ Universit\'e de Toulouse, INSA-CNRS-UPS, LPCNO, 135 Av. Rangueil, 31077 Toulouse, France}

\affiliation{$^3$International Center for Materials Nanoarchitectonics, National Institute for Materials Science, 1-1 Namiki, Tsukuba 305-0044, Japan}

\affiliation{$^4$ Research Center for Functional Materials, National Institute for Materials Science, 1-1 Namiki, Tsukuba 305-0044, Japan}


\begin{abstract}
Hyperspectral imaging at cryogenic temperatures is used to investigate exciton and trion propagation in MoSe$_2$ monolayers encapsulated with hexagonal boron nitride (hBN).  Under a tightly focused, continuous-wave laser excitation, the spatial distribution of neutral excitons and charged trions strongly differ at high excitation densities. Remarkably, in this regime the trion distribution develops a halo shape, similar to that previously observed   in WS$_2$ monolayers at room temperature and under pulsed excitation. In contrast,  the exciton distribution only presents a moderate broadening  without the appereance of a halo. Spatially and spectrally resolved luminescence spectra reveal the buildup of a significant temperature gradient at high excitation power, that is attributed to the energy  relaxation of photoinduced hot carriers. 
 We show, via a numerical resolution of the transport equations for excitons and trions, that the halo can be interpreted  as thermal drift of trions due to a Seebeck term in the particle current.  The model shows that the difference between trion and exciton profiles is simply understood in terms of the very different lifetimes of these two quasiparticles. 
\end{abstract}


\maketitle
\textit{Introduction.---}
In the last decade, two-dimensional crystals of transition metal dichalcogenides (TMDC) such as MX$_2$ (M=Mo, W; X=S, Se, Te) have emerged as promising, atomically thin building blocks for applications in nanoelectronics and optoelectronics \cite{Butler:2013a, Wang:2012c}, as well as ideal platforms for the study of Coulomb-correlated excitations and many-body physics \cite{Chen:2018a,Xiao:2021a,Ugeda:2014a}. Due to quantum confinement, weak dielectric screening and large carrier effective masses in these 2D materials, excitons are characterized by large binding energies and are stable even at room temperature, completely dominating the optical properties of these semiconductors \cite{He:2014a,Chernikov:2014a,Ye:2014a,Qiu:2013a,Ramasubramaniam:2012a, Tuan:2018a}.
 In addition, the interplay between inversion symmetry breaking in monolayers (ML) and the strong spin-orbit coupling induced by the heavy transition metal atoms yields a unique spin/valley locking which is expected to provide additional functionalities in future devices \cite{Xiao:2012a,Li:2020a,Li:2020b,Huang:2020a}.
 Therefore, a better understanding of the transport of excitonic particles is required for the development of future technologies based on the excitonic properties of TMDC MLs and their heterostructures \cite{Unichek:2018a}.  \\
 
\noindent
Very recently, room temperature experiments investigating the temporal and spatial dynamics of the excitonic photoluminescence (PL) under pulsed laser excitation have reported the appereance of a halo in the PL's spatial distribution at high fluence \cite{Kulig:2018a}. This has been attributed to the drift of excitons driven by an effective temperature gradient \cite{Perea:2019a}. Indeed, a strong and focused laser excitation generates  an inhomogeneous distribution of hot carriers. Their subsequent relaxation is at the origin of an inhomogeneous lattice and carrier temperature which in turns pushes excitons out of the excitation spot via the Seebeck effect.  
In this work, we simultaneously measure  the spatial dynamics of excitons and trions in ML MoSe$_2$ using hyperspectral imaging at cryogenic temperatures. Thanks to the encapsulation by  hexagonal boron nitride (hBN), the linewidth of both exciton and trion peaks are close to the homogeneous limit at low temperatures  \cite{Cadiz:2017a}, thus allowing for the clear separation of both excitonic complexes and the study of their propagation in the presence of a common temperature gradient. We show that the thermal drift is 
responsible for the appereance of a halo on the trion's spatial distribution, which is not visible under the exact same conditions for the exciton distribution due to its much shorter lifetime.\\

\begin{figure*}
\includegraphics[width=0.85\textwidth]{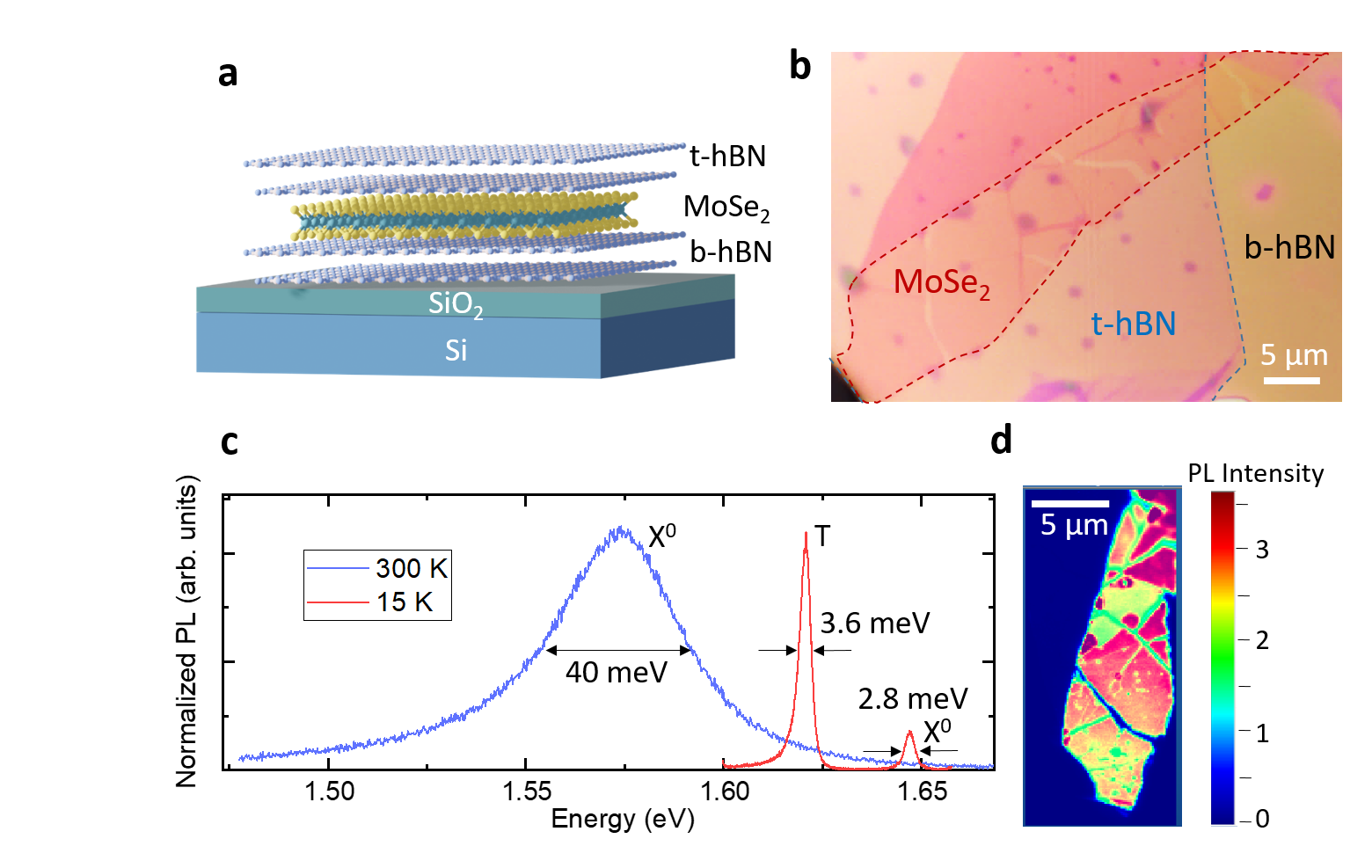}
\caption{\label{fig:fig1} (a) Schematic drawing of the sample. The MoSe$_2$ monolayer is encapsulated between two thin h-BN flakes to provide high optical and transport quality. Here, t-hBN and b-hBN stands for top and bottom hBN, respectively. The whole heterostructure is deposited onto a silicon substrate with a 80 nm oxyde layer. (b) Optical microscope image of the sample. (c)   Normalized photoluminescence spectrum at $T=300$ K (blue) and $T=15$ K (red) under a 1.96 eV cw excitation. The full width at half maximum (FWHM) of each peak is indicated.(d) Spatial map of the total PL intensity at room temperature. Despite bubbles and cracks, one can identify zones which are homogeneous over several $\mu$m.}
\end{figure*}

\indent \textit{Samples and Experimental Set-up.---}
Van der Waals heterostructures such as the one shown in Fig.\ref{fig:fig1} (a) were fabricated by mechanical exfoliation of bulk MoSe$_2$ crystals from 2D semiconductors.  First, a 250 nm thick bottom layer of hBN was mechanically exfoliated onto a freshly cleaved SiO$_2$ (80 nm)/Si substrate using a viscoelastic stamp \cite{Gomez:2014a}. The deposition of the subsequent MoSe$_2$ ML and the second hBN capping layer (thickness 8 nm) is obtained by repeating this procedure. The thicknesses of all the layers were measured by Atomic Force Microscopy (AFM). \\

\noindent
A hyperspectral micro-PL ($\mu$PL) set-up is used to spatially monitor the exciton and trion concentration at cryogenic temperatures \cite{Favorskiy:2010a, Cadiz:2018a}. The samples are kept inside a closed cycle He cryostat and excited either with a red He-Ne laser (1.96 eV) or with  continuous wave tunable lasers. 
Unless otherwise stated, the wavelength of the laser is 570 nm and the sample temperature is $20$ K. The laser beam is focused onto a diffraction-limited spot in the sample plane thanks to a vacuum (and cryogenic) compatible, apochromatic objective mounted inside the cryostat.
  The resulting PL spot is imaged onto the entrance slit of a $320$ mm focal length spectrometer 
 equipped with a 600 grooves/mm diffraction grating. Two methods are used to study the resulting spatial distribution of excitonic complexes: with the slit open, the 0th order specular reflection off the grating is used to image the PL spot onto a cooled Si-CCD camera. Tunable long and short-pass filters 
 were used to selectively image the neutral exciton or the trion's luminesence. This provides, for each line, a spectrally averaged spatial resolution.
The second method consists in closing the entrance slit of the spectrometer so that its image on the CCD is only one pixel wide, and an image of the first order diffraction is recorded. In this configuration, the spatial distribution of the PL is kept in the direction parallel to the slit, whereas the spectral content at a fixed position is found by taking a cross section along the direction perpendicular to the slit.\\

\indent \textit{Results and Discussion.---}
Fig.\ref{fig:fig1} (a) shows a schematic drawing of the Van der Waals heterostructure deposited onto a SiO$_2$/Si substrate. A microscope image of the sample studied here is shown in Fig.\ref{fig:fig1}(b). The PL spectrum under a  $1.96$ eV He-Ne laser excitation at room and cryogenic temperatures is shown in Fig.\ref{fig:fig1}(c). At low temperature, the spectrum exhibits two well defined peaks, ascribed to the neutral exciton ($X^0$) and the charged trion ($T$) which lies  $26.3$ meV below the neutral exciton energy. The trion forms due to  unintentional doping of the MoSe$_2$ monolayer. A spatial map of the total PL intensity at $300$ K is shown in Fig.\ref{fig:fig1}(d), exhibiting regions with uniform PL across several $\mu$m, which are therefore suitable for the study of exciton propagation.\\

\begin{figure*}[htbp]
\includegraphics[width=0.95\textwidth]{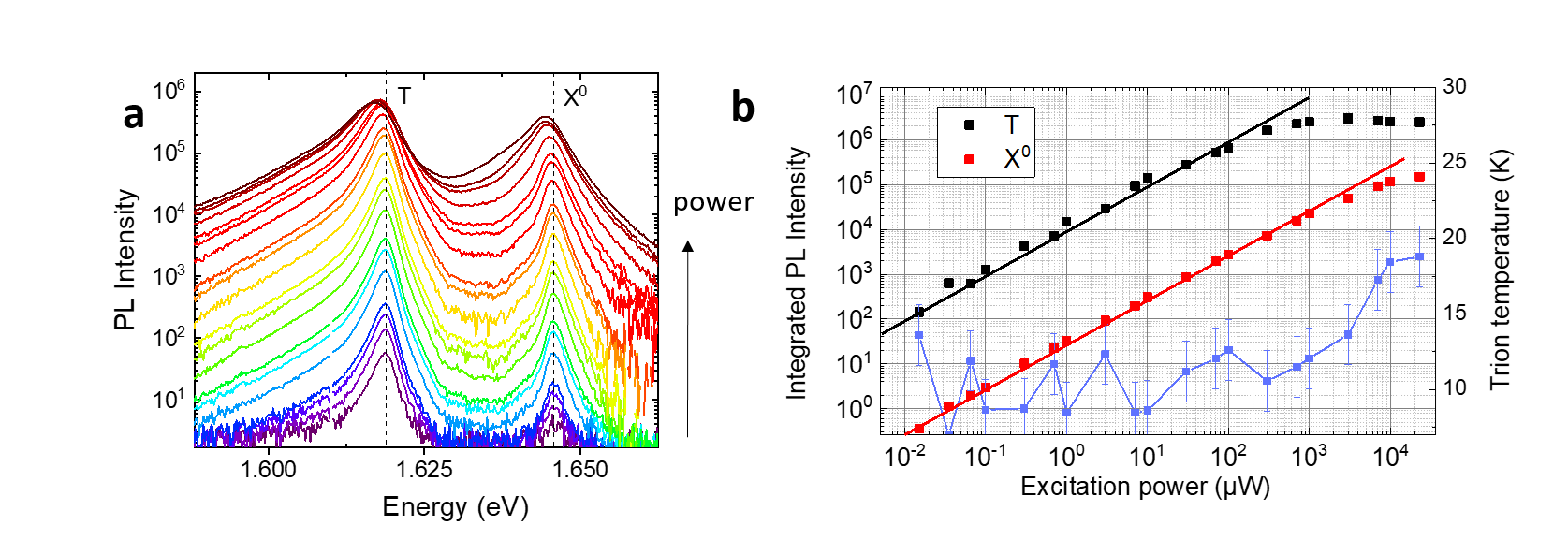}
\caption{\label{fig:fig2} (a) PL spectra for different excitation powers with a continuous wave laser at 570 nm. (b)   Integrated PL intensity for trion and exciton peak as a function of excitation power. The lines correspond to a linear relationship between intensity and excitation power. Also shown is the temperature extracted from the low energy tail of the trion peak, here called Trion temperature.}
\end{figure*}

\noindent
We now focus on the power dependence of the PL properties. The excitation power is varied  over  more than 6 orders of magnitude, from tens of nW up to several tens of mW focused onto a diffraction limited spot, going well beyond the usual power densities used in $\mu$PL experiments. Interestingly, the sample exhibits no evidence of damage after exposure to such power densities, and no irreversible changes in the PL spectrum were observed except for the very first power cycle (see supplementary material), suggesting relatively small photodoping effects which are strongly reduced compared to samples deposited onto SiO$_2$ thanks to the hBN encapsulation \cite{Cadiz:2016b}. Figure \ref{fig:fig2}(a) shows the spatially-averaged PL spectrum on a logarithmic scale for excitation powers ranging from 15 nW up to 23 mW. At high powers, the linewidth of both lines increases while the peak position redshifts at 20 mW by 2 meV and 1 meV for $X^0$ and $T$, respectively. These are typical signatures of local heating of the lattice induced by the optical excitation. Such a  redshift of the neutral exciton would correspond to a heating of the lattice up to $\sim 55$ K  according to the temperature dependence of the PL's peak position (see Supplementary material) However, here the laser excitation at high power may locally change by different amounts both the  lattice and the carrier temperatures.\\

\noindent
In Fig.\ref{fig:fig2}(b) we show the integrated intensity for both T and $X^0$ as a function of excitation power. The continuous lines represent a perfect linear relationship, which is therefore well verified below 100 $\mu$W for both peaks. Remarkably, the trion's PL intensity starts to saturate above 100 $\mu$W and achieves a plateau above  800 $\mu$W , which we attribute to  the fact that the  photoexcitation density becomes larger than the equilibrium carrier density present due to unintentional doping. By taking a trion lifetime of $140$ ps  \cite{Fang:2019a} and an absorption coefficient of $1.5 \times 10^{-3}$ (see supplementary material) we can explain the saturation of the trion's PL  with a resident carrier density of $\sim 5\times 10^{10}\;\mbox{cm}^{-2}$. \\

\noindent
In contrast, the $X^0$ intensity only slightly becomes sublinear above 1 mW, possibly due to Auger-like exciton-exciton annihilation, which has otherwise been shown to be dramatically supressed in hBN encapsulated monolayers. By using a relatively large Auger coefficient of $\gamma \sim 0.1 \; \mbox{cm}^{2}/\mbox{s}$ \cite{Hoshi:2017a,Zipfel:2020a,Cordovilla:2019a}, the absorption coefficient mentioned above and neglecting diffusion (or thermal drift), we indeed expect exciton-exciton annihilation effects to become relevant only above $1$ mW for trions and above $10$ mW for neutral excitons, which is in agreement with the observed behaviour of the $X^0$ PL intensity.\\

\noindent
 Also shown in Fig.\ref{fig:fig2}(b) is the effective temperature of the trion gas extracted from the low energy tail of its luminescence peak, which takes into account  the trion's center-of mass momentum dependence of the transition energy \cite{Christopher:2017a,Zhumagulov:2020a}. The trion's luminescence intensity is :

\begin{equation}
I(h \nu) \propto e^{-(E^0-h\nu)/\varepsilon} \Theta\left(E^0-h\nu\right) \ast g(h\nu)
\label{eq1}
\end{equation}

\noindent
where $h\nu$ is the emitted photon's energy, $E_0$ is the emission energy of a trion with zero center-of-mass momentum,  $g$ is a Lorentzian representing the non-zero linewidth of the transition, $\Theta$ is the Heaviside function, and $\varepsilon \approx 2 k_b T$ where $k_B$ is Boltzmann's constant. The latter approximation is valid since  the trion's radius is of the order of nanometers and  the electron and hole effecive masses are comparable. As seen in Fig.\ref{fig:fig2}(b) (blue line and markers), the spatially averaged trion temperature significantly increases above 1 mW.   As will be shown later, this spatial averaging strongly underestimates the increase of the effective temperature due to the peculiar spatial distribution of trions at high densities. Note that the trion temperature at low excitation, on the order of $10$ K, is smaller than the lattice temperature. This trend is verified at different lattice temperatures, as shown in the Supplementary material. We cannot exclude a systematic error present in the model behind Eq.(\ref{eq1}) whose precise determination goes beyond the scope of this work. \\

\begin{figure*}[htbp]
\includegraphics[width=0.95\textwidth]{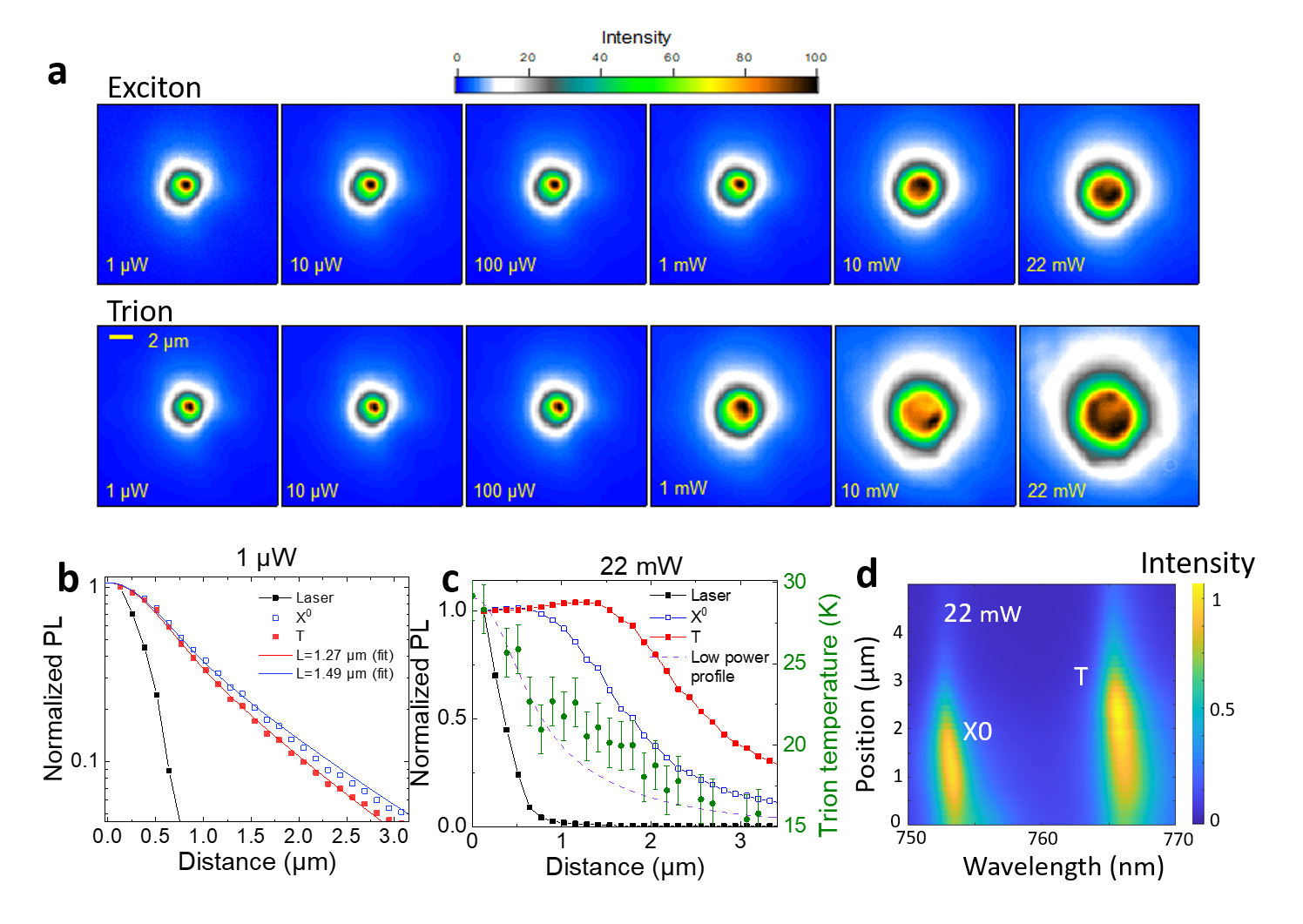}
\caption{\label{fig:fig3} (a) PL's spatial distribution of excitons and trions for selected excitation powers. The images have a lateral size of $14 \;\mu$m. (b) Normalized radial profile for excitons and trions at low excitation power in a logarithmic scale. Also shown is the laser profile and fits of the 2D diffusion equation. (c) Radial profiles at the highest excitation power, normalized to one at the center in a linear scale. Also shown is the extracted trion temperature from the image of panel (d). The dashed line corresponds to the exciton profile at low power.(d) First order diffraction image at high excitation power, allowing to obtain the spectrum as a function of the distance to the excitation spot. 
}
\end{figure*}

\noindent
In order to study the effect of this local heating on excitonic transport, we performed spatially resolved microluminescence imaging. The  diffraction-limited Airy disk of the laser in the sample plane can  be approximated at short distances by a Gaussian profile of the form $e^{-r^2/w^2}$, where $w \approx 0.5 \;\mu m$. Figure \ref{fig:fig3}(a) shows the spatial images of the luminescence of both excitons and trions, each properly filtered by interferential tunable filters. Remarkably, the spatial distribution of excitons and trions are practically equal at low power, but  significantly differ at high excitation, due to the appearance of a halo in the trion's intensity at the center, similar to the halo observed at 300 K on WS$_2$ monolayers \cite{Kulig:2018a,Perea:2019a}, with the main difference here being the cryogenic temperature and a continuous  excitation.\\

\noindent
 Due to the rotational symmetry of the PL images,  one-dimensional profiles are obtained by averaging the PL intensity along different radial directions. The result is shown in Fig. \ref{fig:fig3}(b) for an excitation power of $1 \; \mu$W, where it can be seen that the exciton and trion luminescence comes from a spatial region which extends significantly beyond the excitation spot. Suprisingly, both profiles are very similar. A fit with the analytical solution of the 2D steady-state diffusion equation \cite{Cadiz:2017a} yields similar effective diffusion lengths, of $1.49 \pm 0.05\;\mu$m and $1.27\pm 0.05\;\mu$m for $X^0$ and $T$, respectively. This result is already surprising given that the lifetime of both species differ by more than one order of magnitude. As suggested by recent experiments  in encapsulated MoSe$_2$ monolayers \cite{Fang:2019a}, we interpret these low-excitation profiles as being the result of diffusion of hot excitons, which relax into bright excitons and trions on a timescale of the order of $18$ ps. If the subsequent diffusion of excitons and trions is slow enough, then we would indeed expect to see the same profile for both species. Note that momentum-indirect dark excitons can also constitute a reservoir of excitons at the steady state and possibly contribute to the low excitation profile  \cite{Madeo:2020a}. \\

\noindent
Now we discuss the high power regime, characterized by the spatial broadening of the PL profiles and the appereance of a dip at the center of the trion's spatial distribution, clearly visible in the radial profile of Fig. \ref{fig:fig3}(c). We interpret this as a thermal drift induced by a temperature gradient. Indeed, the measured trion's temperature as a function of space is also shown and exhibits a monotonic decrease from $30$ K at the center to $15$ K at a few microns.   This was obtained by recording the 
1st order diffraction image on the CCD detector, shown in Fig. \ref{fig:fig3}(d). We attribute the formation of a temperature gradient as a consequence of the energy relaxation of hot carriers by emission of phonons. This gives rise to a hot spot and subsequent thermal drift of excitons and trions. This scenario is in agreement with the fact that the observed temperature gradient exhibits a similar radial profile as the low excitation power profiles observed for $X^0$ and $T$, previously attributed to the hot carrier distribution. Note that in these encapsulated monolayers, Auger recombination does not seem to play a significant role in the halo formation under non-resonant excitation, in contrast with the situation found in non encapsulated samples at room temperature \cite{Perea:2019a}. Indeed, if the temperature gradient was created by Auger-like non radiative recombination, we would expect the steady-state spatial dependence of the temperature to be non monotonic and to exhibit a halo as well.\\

\noindent
To justify this picture, we have modeled the exciton and trion's profiles by numerically solving the 2D steady state drift-diffusion equation:

\begin{equation}
G - \frac{n}{\tau} - \gamma n^2 = \vec \nabla \cdot \left (  - D \vec \nabla n - \sigma S  \vec \nabla T  \right) 
 \label{eq2}
\end{equation}

\noindent
where $n$ is the concentration of either $X^0$ and $T$, $G$ is the generation rate, here representing the spatial distribution of hot carriers (low excitation profiles in Fig. \ref{fig:fig3}(b)), $\tau$ is the lifetime, $\gamma$ is the Auger coefficient, $D= (k_BT/q) \mu $  with $\mu$ the mobility, $q$ the elementary charge, $\sigma = q \mu n$ is the conductivity and $S$ the Seebeck coefficient, which is of the order of $S\sim 2 k_B/q$. Details of this calculation are shown in the supplementary material. Note that the Seebeck coefficient $S$, which is here a universal constant  independent on doping and temperature, is not the same as the phenomenological Seebeck coefficient usually measured in thermoelectric experiments which describes the proportionality between the electrochemical potential gradient (and not the electric field) and the temperature gradient \cite{Cai:2006a}.  In addition, in this work we suppose for the sake of simplicity that trions can be modelled by a unipolar diffusion equation. In reality, charge conservation implies that trions cannot diffuse unless the background charge density is modified accordingly. If the trion density remains small compared to the background resident carrier density, which is actually guaranteed by the saturation of the trion formation rate, we may employ a unipolar model. Finally, we also suppose that there is no coupling between the trion and exciton populations, which is in agreement to previous time-resolved measurements where totally independent dynamics has been observed for both complexes.\\

 \begin{figure*}[htbp]
\includegraphics[width=0.75\textwidth]{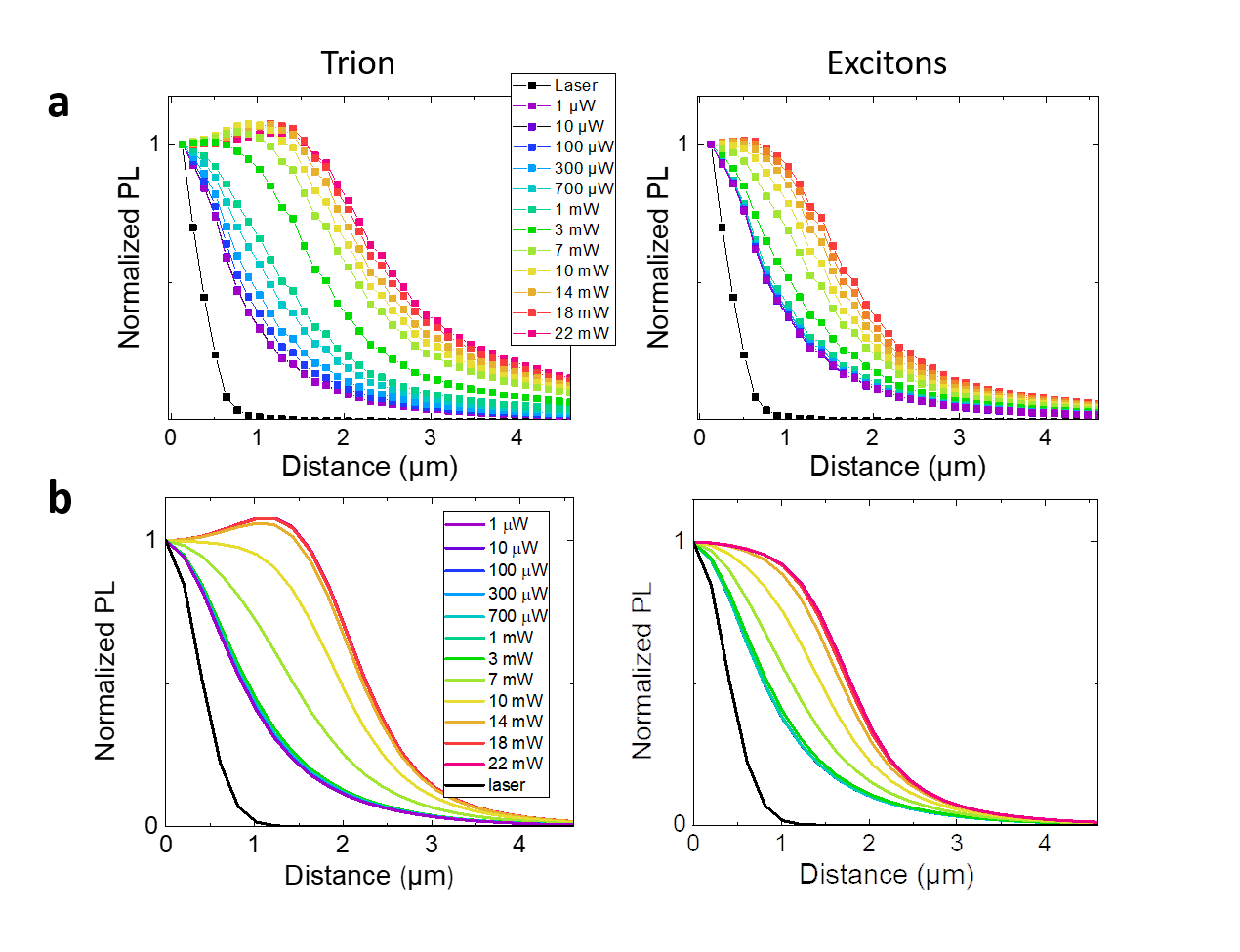}
\caption{\label{fig:fig4} (a) Measured PL's radial profile for trions (left) and excitons (right) for selected excitation powers.  (b) Calculated profiles obtained with a numerical solution of Eq.(\ref{eq2})
}
\end{figure*}

\noindent
Figure \ref{fig:fig4} shows a comparison between the measured radial profiles and the ones predicted by the numerical solution of Eq. (\ref{eq2}) when taking a mobility of $\mu = 100 \; \mbox{cm}^2V^{-1}\mbox{s}^{-1}$ for  trions and  $\mu = 150 \; \mbox{cm}^2V^{-1}\mbox{s}^{-1}$  for excitons. The Seebeck coefficients were chosen as $S=300 \;\mu$V/K and  $S=400 \;\mu$V/K  for trions and excitons, respectively. The trion lifetime of $\tau=140$ ps is extracted from \cite{Fang:2019a} and the exciton lifetime is estimated to be $\tau=5$ ps by considering an intrinsic radiative lifetime (in vacuum) of $2.7$ ps and cavity effects.
The model reproduces well the main features of both the exciton and trion profiles with a reasonable choice of adjustable parameters.  At this stage, we cannot claim a precise measurement of the Seebeck coefficient, since for example the mobility $\mu$ is unknown and we do not claim a precise determination of the absolute trion temperature. We have chosen  $\mu = 100 \; \mbox{cm}^2V^{-1}s^{-1}$ for the trion  which is low enough to ensure a low power profile dominated by hot exciton diffusion. For the excitons, we have taken into account the effective mass difference, although little is known about the momentum relaxation time for both species.  \\

\noindent
Note that the Seebeck coefficients for excitons and trions do not need to be equal. Indeed, for a classical 2D gas for which the momentum relaxation time $\tau$ depends on the kinetic energy $\varepsilon$ as $\tau (\varepsilon) \propto \epsilon^r $, the Seebeck coefficient writes $S= (2+r)k_B/q$ and $r$ can differ between excitons and trions.\\ 

\begin{figure*}[htbp]
\includegraphics[width=0.95\textwidth]{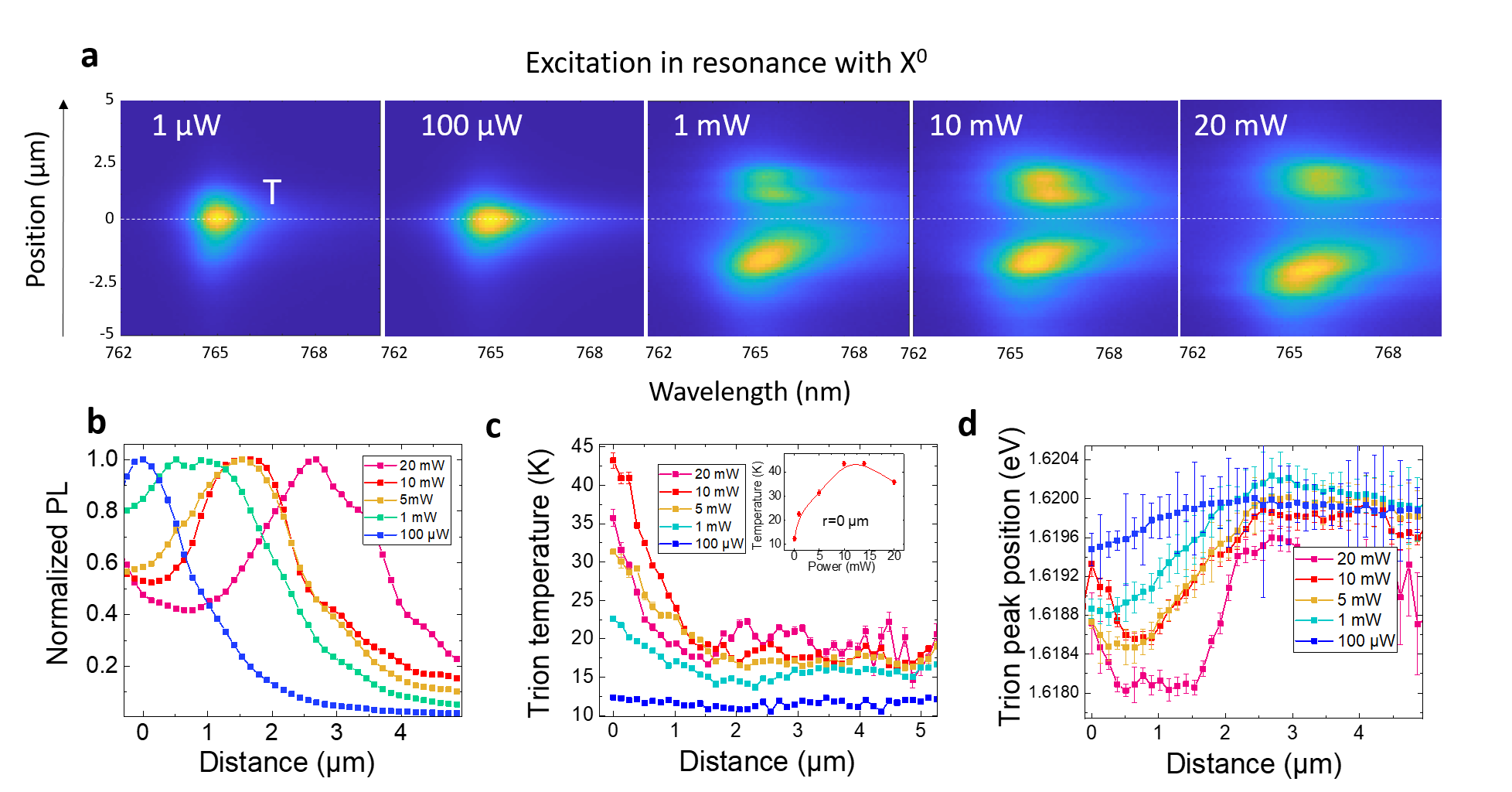}
\caption{\label{fig:fig5} (a) Spatial dependence of the trion's luminesence spectrum for selected excitation powers when exciting resonantly the neutral exciton (b) Normalized trion profiles obtained from the figures in (a) by integrating the trion's spectrum at different distances from the excitation spot. (c) Trion's temperature as a function of position. The inset shows the measured temperature at the excitation spot. (d) Trion's peak energy as a function of position.
}
\end{figure*}

\noindent
Finally, we show that the trion's halo formation can be further enhanced with resonance excitation. This is in contrast with recent experiments performed on high purity GaAs samples, for which  halo formation of excitons was attributed to a spatially dependent population balance between excitons and free electron hole plasma, and shown to disappear under resonance excitation \cite{Bieker:2015a}. 
 Here, the laser energy is set to match the $X^0$ energy. The absorption, of around $5 \;\%$, is estimated by considering optical cavity effects.
 Fig. \ref{fig:fig5}(a) shows the obtained first order diffraction images at different excitation powers. At high powers, the dip in the trion's luminescence becomes very pronounced, with a maximum trion intensity at distances lager than $2\;\mu$m from the center at 20 mW, as shown in the radial profiles in Fig. \ref{fig:fig5}(b). The spatial dependence of the trion's temperature is shown in Fig. \ref{fig:fig5} (c), where it can be seen that the temperature gradient is a monotonic function of the excitation power below 10 mW. The fact that a temperature gradient exists despite resonant excitation could indicate an important role played by resonant exciton-exciton Auger scattering involving an excited conduction band, which was recently proposed to explain the very efficient PL upconversion observed  in MoTe$_2$ and MoSe$_2$ MLs under resonant excitation of the neutral exciton \cite{Han:2018a}.

  The inset of Fig. \ref{fig:fig5}(c) shows that the temperature at the excitation spot starts to saturate above 10 mW and even decreases at 20 mW. Note that a saturation of the excess temperature of a hot exciton gas at high densities has been previously reported in encapsulated WSe$_2$ MLs and attributed to the balancing between Auger heating and phonon cooling \cite{Cordovilla:2019a}. In addition, due to the resonant excitation condition, at the highest excitation powers the redshift of the exciton transition may result in a lower absorption coefficient, counterbalancing the effect of a higher excitation on the photocreated density. Fig. \ref{fig:fig5} (d) shows the trion's peak energy as a function of space. Above 5 mW, a non-monotonic dependence is observed, with a remarkable blueshift-redshift crossover  near the center. Note that such a  crossover of the exciton absorption energy on ML WS$_2$ at 300 K has been reported and attributed to the onset of a repulsive exciton-exciton interaction at the Mott density $2 \times 10^{12}\;\mbox{cm}^{-2}$\cite{Sie:2017a}.  Here, the crossover can be equally attributed to a repulsive interaction between hot excitons which, near the center, are then converted into trions by binding to an extra charge carrier.\\

\indent \textit{Conclusions}
In conclusion, we have imaged the spatial distribution of trions and excitons in ML MoSe$_2$ at cryogenic temperatures as a function of excitation power. We reveal the appereance of a halo in the trion's distribution, which so far was only observed under pulsed laser excitation at room temperature. Hyperspectral imaging permits a measurement of the trion temperature as a function of space and confirms the existence of a strong temperature gradient at high excitation. The results are consistent with the following simple picture: photons generate hot excitons out of the light cone, which then diffuse over typically one micron before relaxing their energy and convert into either bright excitons or trions. This energy relaxation creates a temperature gradient which extends across the hot carrier distribution. Excitons and trions then drift under the presence of this temperature  gradient, which dominates over diffusion. The measured exciton profiles, which do not exhibit a halo but a moderate broadening, are consistent with the exciton having a much shorter lifetime. Finally, we have shown that this halo can be observed under resonant excitation of the neutral exciton, which can be explained by the fact that resonant exciton-exciton interactions  creates hot excitons, thus mimicking the effect of an off-resonance excitation. This work is a step towards a more detailed understanding of the different physical phenomena that can govern excitonic transport at high excitation fluences.\\

\indent \textit{Acknowledgements.---} 
F.C, S. P, A. R, D.P and F. S ackowledge the Grant "SpinCAT" No. ANR-18-CE24-0011-01 
. X.M. also acknowledges the Institut Universitaire de France. C.R.  acknowledges ANR Vallex and Labex NEXT.  K.W. and T.T. acknowledge support from the Elemental Strategy Initiative
conducted by the MEXT, Japan ,Grant Number JPMXP0112101001 and JSPS
KAKENHI Grant Number JP20H00354.

\bibliography{2dmatforTrion}

\begin{thebibliography}{34}%
\makeatletter
\providecommand \@ifxundefined [1]{%
 \@ifx{#1\undefined}
}%
\providecommand \@ifnum [1]{%
 \ifnum #1\expandafter \@firstoftwo
 \else \expandafter \@secondoftwo
 \fi
}%
\providecommand \@ifx [1]{%
 \ifx #1\expandafter \@firstoftwo
 \else \expandafter \@secondoftwo
 \fi
}%
\providecommand \natexlab [1]{#1}%
\providecommand \enquote  [1]{``#1''}%
\providecommand \bibnamefont  [1]{#1}%
\providecommand \bibfnamefont [1]{#1}%
\providecommand \citenamefont [1]{#1}%
\providecommand \href@noop [0]{\@secondoftwo}%
\providecommand \href [0]{\begingroup \@sanitize@url \@href}%
\providecommand \@href[1]{\@@startlink{#1}\@@href}%
\providecommand \@@href[1]{\endgroup#1\@@endlink}%
\providecommand \@sanitize@url [0]{\catcode `\\12\catcode `\$12\catcode
  `\&12\catcode `\#12\catcode `\^12\catcode `\_12\catcode `\%12\relax}%
\providecommand \@@startlink[1]{}%
\providecommand \@@endlink[0]{}%
\providecommand \url  [0]{\begingroup\@sanitize@url \@url }%
\providecommand \@url [1]{\endgroup\@href {#1}{\urlprefix }}%
\providecommand \urlprefix  [0]{URL }%
\providecommand \Eprint [0]{\href }%
\providecommand \doibase [0]{https://doi.org/}%
\providecommand \selectlanguage [0]{\@gobble}%
\providecommand \bibinfo  [0]{\@secondoftwo}%
\providecommand \bibfield  [0]{\@secondoftwo}%
\providecommand \translation [1]{[#1]}%
\providecommand \BibitemOpen [0]{}%
\providecommand \bibitemStop [0]{}%
\providecommand \bibitemNoStop [0]{.\EOS\space}%
\providecommand \EOS [0]{\spacefactor3000\relax}%
\providecommand \BibitemShut  [1]{\csname bibitem#1\endcsname}%
\let\auto@bib@innerbib\@empty
\bibitem [{\citenamefont {Butler}\ \emph {et~al.}(2013)\citenamefont {Butler},
  \citenamefont {Hollen}, \citenamefont {Cao}, \citenamefont {Cui},
  \citenamefont {Gupta}, \citenamefont {Guti{\'e}rrez}, \citenamefont {Heinz},
  \citenamefont {Hong}, \citenamefont {Huang}, \citenamefont {Ismach},
  \citenamefont {Johnston-Halperin}, \citenamefont {Kuno}, \citenamefont
  {Plashnitsa}, \citenamefont {Robinson}, \citenamefont {Ruoff}, \citenamefont
  {Salahuddin}, \citenamefont {Shan}, \citenamefont {Shi}, \citenamefont
  {Spencer}, \citenamefont {Terrones}, \citenamefont {Windl},\ and\
  \citenamefont {Goldberger}}]{Butler:2013a}%
  \BibitemOpen
  \bibfield  {author} {\bibinfo {author} {\bibfnamefont {S.~Z.}\ \bibnamefont
  {Butler}}, \bibinfo {author} {\bibfnamefont {S.~M.}\ \bibnamefont {Hollen}},
  \bibinfo {author} {\bibfnamefont {L.}~\bibnamefont {Cao}}, \bibinfo {author}
  {\bibfnamefont {Y.}~\bibnamefont {Cui}}, \bibinfo {author} {\bibfnamefont
  {J.~A.}\ \bibnamefont {Gupta}}, \bibinfo {author} {\bibfnamefont {H.~R.}\
  \bibnamefont {Guti{\'e}rrez}}, \bibinfo {author} {\bibfnamefont {T.~F.}\
  \bibnamefont {Heinz}}, \bibinfo {author} {\bibfnamefont {S.~S.}\ \bibnamefont
  {Hong}}, \bibinfo {author} {\bibfnamefont {J.}~\bibnamefont {Huang}},
  \bibinfo {author} {\bibfnamefont {A.~F.}\ \bibnamefont {Ismach}}, \bibinfo
  {author} {\bibfnamefont {E.}~\bibnamefont {Johnston-Halperin}}, \bibinfo
  {author} {\bibfnamefont {M.}~\bibnamefont {Kuno}}, \bibinfo {author}
  {\bibfnamefont {V.~V.}\ \bibnamefont {Plashnitsa}}, \bibinfo {author}
  {\bibfnamefont {R.~D.}\ \bibnamefont {Robinson}}, \bibinfo {author}
  {\bibfnamefont {R.~S.}\ \bibnamefont {Ruoff}}, \bibinfo {author}
  {\bibfnamefont {S.}~\bibnamefont {Salahuddin}}, \bibinfo {author}
  {\bibfnamefont {J.}~\bibnamefont {Shan}}, \bibinfo {author} {\bibfnamefont
  {L.}~\bibnamefont {Shi}}, \bibinfo {author} {\bibfnamefont {M.~G.}\
  \bibnamefont {Spencer}}, \bibinfo {author} {\bibfnamefont {M.}~\bibnamefont
  {Terrones}}, \bibinfo {author} {\bibfnamefont {W.}~\bibnamefont {Windl}},\
  and\ \bibinfo {author} {\bibfnamefont {J.~E.}\ \bibnamefont {Goldberger}},\
  }\bibfield  {title} {\bibinfo {title} {Progress, challenges, and
  opportunities in two-dimensional materials beyond graphene},\ }\href@noop {}
  {\bibfield  {journal} {\bibinfo  {journal} {ACS Nano}\ }\textbf {\bibinfo
  {volume} {7}},\ \bibinfo {pages} {2898} (\bibinfo {year} {2013})}\BibitemShut
  {NoStop}%
\bibitem [{\citenamefont {Wang}\ \emph {et~al.}(2012)\citenamefont {Wang},
  \citenamefont {Kalantar-Zadeh}, \citenamefont {Kis}, \citenamefont
  {Coleman},\ and\ \citenamefont {Strano}}]{Wang:2012c}%
  \BibitemOpen
  \bibfield  {author} {\bibinfo {author} {\bibfnamefont {Q.~H.}\ \bibnamefont
  {Wang}}, \bibinfo {author} {\bibfnamefont {K.}~\bibnamefont
  {Kalantar-Zadeh}}, \bibinfo {author} {\bibfnamefont {A.}~\bibnamefont {Kis}},
  \bibinfo {author} {\bibfnamefont {J.~N.}\ \bibnamefont {Coleman}},\ and\
  \bibinfo {author} {\bibfnamefont {M.~S.}\ \bibnamefont {Strano}},\ }\bibfield
   {title} {\bibinfo {title} {Electronics and optoelectronics of
  two-dimensional transition metal dichalcogenides},\ }\href@noop {} {\bibfield
   {journal} {\bibinfo  {journal} {Nature nanotechnology}\ }\textbf {\bibinfo
  {volume} {7}},\ \bibinfo {pages} {699} (\bibinfo {year} {2012})}\BibitemShut
  {NoStop}%
\bibitem [{\citenamefont {Chen}\ \emph {et~al.}(2018)\citenamefont {Chen},
  \citenamefont {Goldstein}, \citenamefont {Taniguchi}, \citenamefont
  {Watanabe},\ and\ \citenamefont {Yan}}]{Chen:2018a}%
  \BibitemOpen
  \bibfield  {author} {\bibinfo {author} {\bibfnamefont {S.~Y.}\ \bibnamefont
  {Chen}}, \bibinfo {author} {\bibfnamefont {T.}~\bibnamefont {Goldstein}},
  \bibinfo {author} {\bibfnamefont {T.}~\bibnamefont {Taniguchi}}, \bibinfo
  {author} {\bibfnamefont {K.}~\bibnamefont {Watanabe}},\ and\ \bibinfo
  {author} {\bibfnamefont {J.}~\bibnamefont {Yan}},\ }\bibfield  {title}
  {\bibinfo {title} {Coulomb-bound four- and five-particle intervalley states
  in an atomically-thin semiconductor},\ }\href@noop {} {\bibfield  {journal}
  {\bibinfo  {journal} {Nature Communications}\ }\textbf {\bibinfo {volume}
  {9}},\ \bibinfo {pages} {3717} (\bibinfo {year} {2018})}\BibitemShut
  {NoStop}%
\bibitem [{\citenamefont {Xiao}\ \emph {et~al.}(2021)\citenamefont {Xiao},
  \citenamefont {Yan}, \citenamefont {Liu}, \citenamefont {Yang}, \citenamefont
  {Kan}, \citenamefont {Duan}, \citenamefont {Liu},\ and\ \citenamefont
  {Cui}}]{Xiao:2021a}%
  \BibitemOpen
  \bibfield  {author} {\bibinfo {author} {\bibfnamefont {K.}~\bibnamefont
  {Xiao}}, \bibinfo {author} {\bibfnamefont {T.}~\bibnamefont {Yan}}, \bibinfo
  {author} {\bibfnamefont {Q.}~\bibnamefont {Liu}}, \bibinfo {author}
  {\bibfnamefont {S.}~\bibnamefont {Yang}}, \bibinfo {author} {\bibfnamefont
  {C.}~\bibnamefont {Kan}}, \bibinfo {author} {\bibfnamefont {R.}~\bibnamefont
  {Duan}}, \bibinfo {author} {\bibfnamefont {Z.}~\bibnamefont {Liu}},\ and\
  \bibinfo {author} {\bibfnamefont {X.}~\bibnamefont {Cui}},\ }\bibfield
  {title} {\bibinfo {title} {Many-body effect on optical properties of
  monolayer molybdenum diselenide},\ }\href@noop {} {\bibfield  {journal}
  {\bibinfo  {journal} {J. Phys. Chem. Lett}\ }\textbf {\bibinfo {volume}
  {12}},\ \bibinfo {pages} {2555} (\bibinfo {year} {2021})}\BibitemShut
  {NoStop}%
\bibitem [{\citenamefont {{Ugeda}}\ \emph {et~al.}(2014)\citenamefont
  {{Ugeda}}, \citenamefont {{Bradley}}, \citenamefont {{Shi}}, \citenamefont
  {{da Jornada}}, \citenamefont {{Zhang}}, \citenamefont {{Qiu}}, \citenamefont
  {{Mo}}, \citenamefont {{Hussain}}, \citenamefont {{Shen}}, \citenamefont
  {{Wang}}, \citenamefont {{Louie}},\ and\ \citenamefont
  {{Crommie}}}]{Ugeda:2014a}%
  \BibitemOpen
  \bibfield  {author} {\bibinfo {author} {\bibfnamefont {M.~M.}\ \bibnamefont
  {{Ugeda}}}, \bibinfo {author} {\bibfnamefont {A.~J.}\ \bibnamefont
  {{Bradley}}}, \bibinfo {author} {\bibfnamefont {S.-F.}\ \bibnamefont
  {{Shi}}}, \bibinfo {author} {\bibfnamefont {F.~H.}\ \bibnamefont {{da
  Jornada}}}, \bibinfo {author} {\bibfnamefont {Y.}~\bibnamefont {{Zhang}}},
  \bibinfo {author} {\bibfnamefont {D.~Y.}\ \bibnamefont {{Qiu}}}, \bibinfo
  {author} {\bibfnamefont {S.-K.}\ \bibnamefont {{Mo}}}, \bibinfo {author}
  {\bibfnamefont {Z.}~\bibnamefont {{Hussain}}}, \bibinfo {author}
  {\bibfnamefont {Z.-X.}\ \bibnamefont {{Shen}}}, \bibinfo {author}
  {\bibfnamefont {F.}~\bibnamefont {{Wang}}}, \bibinfo {author} {\bibfnamefont
  {S.~G.}\ \bibnamefont {{Louie}}},\ and\ \bibinfo {author} {\bibfnamefont
  {M.~F.}\ \bibnamefont {{Crommie}}},\ }\bibfield  {title} {\bibinfo {title}
  {Observation of giant bandgap renormalization and excitonic effects in a
  monolayer transition metal dichalcogenide semiconductor},\ }\href@noop {}
  {\bibfield  {journal} {\bibinfo  {journal} {Nature Materials}\ }\textbf
  {\bibinfo {volume} {13}},\ \bibinfo {pages} {1091} (\bibinfo {year}
  {2014})}\BibitemShut {NoStop}%
\bibitem [{\citenamefont {He}\ \emph {et~al.}(2014)\citenamefont {He},
  \citenamefont {Kumar}, \citenamefont {Zhao}, \citenamefont {Wang},
  \citenamefont {Mak}, \citenamefont {Zhao},\ and\ \citenamefont
  {Shan}}]{He:2014a}%
  \BibitemOpen
  \bibfield  {author} {\bibinfo {author} {\bibfnamefont {K.}~\bibnamefont
  {He}}, \bibinfo {author} {\bibfnamefont {N.}~\bibnamefont {Kumar}}, \bibinfo
  {author} {\bibfnamefont {L.}~\bibnamefont {Zhao}}, \bibinfo {author}
  {\bibfnamefont {Z.}~\bibnamefont {Wang}}, \bibinfo {author} {\bibfnamefont
  {K.~F.}\ \bibnamefont {Mak}}, \bibinfo {author} {\bibfnamefont
  {H.}~\bibnamefont {Zhao}},\ and\ \bibinfo {author} {\bibfnamefont
  {J.}~\bibnamefont {Shan}},\ }\bibfield  {title} {\bibinfo {title} {Tightly
  bound excitons in monolayer ${\mathrm{wse}}_{2}$},\ }\href
  {https://doi.org/10.1103/PhysRevLett.113.026803} {\bibfield  {journal}
  {\bibinfo  {journal} {Phys. Rev. Lett.}\ }\textbf {\bibinfo {volume} {113}},\
  \bibinfo {pages} {026803} (\bibinfo {year} {2014})}\BibitemShut {NoStop}%
\bibitem [{\citenamefont {Chernikov}\ \emph {et~al.}(2014)\citenamefont
  {Chernikov}, \citenamefont {Berkelbach}, \citenamefont {Hill}, \citenamefont
  {Rigosi}, \citenamefont {Li}, \citenamefont {Aslan}, \citenamefont
  {Reichman}, \citenamefont {Hybertsen},\ and\ \citenamefont
  {Heinz}}]{Chernikov:2014a}%
  \BibitemOpen
  \bibfield  {author} {\bibinfo {author} {\bibfnamefont {A.}~\bibnamefont
  {Chernikov}}, \bibinfo {author} {\bibfnamefont {T.~C.}\ \bibnamefont
  {Berkelbach}}, \bibinfo {author} {\bibfnamefont {H.~M.}\ \bibnamefont
  {Hill}}, \bibinfo {author} {\bibfnamefont {A.}~\bibnamefont {Rigosi}},
  \bibinfo {author} {\bibfnamefont {Y.}~\bibnamefont {Li}}, \bibinfo {author}
  {\bibfnamefont {O.~B.}\ \bibnamefont {Aslan}}, \bibinfo {author}
  {\bibfnamefont {D.~R.}\ \bibnamefont {Reichman}}, \bibinfo {author}
  {\bibfnamefont {M.~S.}\ \bibnamefont {Hybertsen}},\ and\ \bibinfo {author}
  {\bibfnamefont {T.~F.}\ \bibnamefont {Heinz}},\ }\bibfield  {title} {\bibinfo
  {title} {Exciton binding energy and nonhydrogenic rydberg series in monolayer
  {W}{S}2},\ }\href {https://doi.org/10.1103/PhysRevLett.113.076802} {\bibfield
   {journal} {\bibinfo  {journal} {Phys. Rev. Lett.}\ }\textbf {\bibinfo
  {volume} {113}},\ \bibinfo {pages} {076802} (\bibinfo {year}
  {2014})}\BibitemShut {NoStop}%
\bibitem [{\citenamefont {{Ye}}\ \emph {et~al.}(2014)\citenamefont {{Ye}},
  \citenamefont {{Cao}}, \citenamefont {{O'Brien}}, \citenamefont {{Zhu}},
  \citenamefont {{Yin}}, \citenamefont {{Wang}}, \citenamefont {{Louie}},\ and\
  \citenamefont {{Zhang}}}]{Ye:2014a}%
  \BibitemOpen
  \bibfield  {author} {\bibinfo {author} {\bibfnamefont {Z.}~\bibnamefont
  {{Ye}}}, \bibinfo {author} {\bibfnamefont {T.}~\bibnamefont {{Cao}}},
  \bibinfo {author} {\bibfnamefont {K.}~\bibnamefont {{O'Brien}}}, \bibinfo
  {author} {\bibfnamefont {H.}~\bibnamefont {{Zhu}}}, \bibinfo {author}
  {\bibfnamefont {X.}~\bibnamefont {{Yin}}}, \bibinfo {author} {\bibfnamefont
  {Y.}~\bibnamefont {{Wang}}}, \bibinfo {author} {\bibfnamefont {S.~G.}\
  \bibnamefont {{Louie}}},\ and\ \bibinfo {author} {\bibfnamefont
  {X.}~\bibnamefont {{Zhang}}},\ }\bibfield  {title} {\bibinfo {title}
  {{Probing Excitonic Dark States in Single-layer Tungsten Disulfide}},\
  }\href@noop {} {\bibfield  {journal} {\bibinfo  {journal} {Nature}\ }\textbf
  {\bibinfo {volume} {513}},\ \bibinfo {pages} {214} (\bibinfo {year}
  {2014})}\BibitemShut {NoStop}%
\bibitem [{\citenamefont {Qiu}\ \emph {et~al.}(2013)\citenamefont {Qiu},
  \citenamefont {da~Jornada},\ and\ \citenamefont {Louie}}]{Qiu:2013a}%
  \BibitemOpen
  \bibfield  {author} {\bibinfo {author} {\bibfnamefont {D.~Y.}\ \bibnamefont
  {Qiu}}, \bibinfo {author} {\bibfnamefont {F.~H.}\ \bibnamefont
  {da~Jornada}},\ and\ \bibinfo {author} {\bibfnamefont {S.~G.}\ \bibnamefont
  {Louie}},\ }\bibfield  {title} {\bibinfo {title} {Optical spectrum of
  {M}o{S}2: Many-body effects and diversity of exciton states},\ }\href@noop {}
  {\bibfield  {journal} {\bibinfo  {journal} {Phys. Rev. Lett.}\ }\textbf
  {\bibinfo {volume} {111}},\ \bibinfo {pages} {216805} (\bibinfo {year}
  {2013})}\BibitemShut {NoStop}%
\bibitem [{\citenamefont {Ramasubramaniam}(2012)}]{Ramasubramaniam:2012a}%
  \BibitemOpen
  \bibfield  {author} {\bibinfo {author} {\bibfnamefont {A.}~\bibnamefont
  {Ramasubramaniam}},\ }\bibfield  {title} {\bibinfo {title} {Large excitonic
  effects in monolayers of molybdenum and tungsten dichalcogenides},\
  }\href@noop {} {\bibfield  {journal} {\bibinfo  {journal} {Phys. Rev. B}\
  }\textbf {\bibinfo {volume} {86}},\ \bibinfo {pages} {115409} (\bibinfo
  {year} {2012})}\BibitemShut {NoStop}%
\bibitem [{\citenamefont {Tuan}\ \emph {et~al.}(2018)\citenamefont {Tuan},
  \citenamefont {Yang},\ and\ \citenamefont {Dery}}]{Tuan:2018a}%
  \BibitemOpen
  \bibfield  {author} {\bibinfo {author} {\bibfnamefont {D.~V.}\ \bibnamefont
  {Tuan}}, \bibinfo {author} {\bibfnamefont {M.}~\bibnamefont {Yang}},\ and\
  \bibinfo {author} {\bibfnamefont {H.}~\bibnamefont {Dery}},\ }\bibfield
  {title} {\bibinfo {title} {Coulomb interaction in monolayer transition-metal
  dichalcogenides},\ }\href@noop {} {\bibfield  {journal} {\bibinfo  {journal}
  {Physical Review B}\ }\textbf {\bibinfo {volume} {98}},\ \bibinfo {pages}
  {125308} (\bibinfo {year} {2018})}\BibitemShut {NoStop}%
\bibitem [{\citenamefont {Xiao}\ \emph {et~al.}(2012)\citenamefont {Xiao},
  \citenamefont {Liu}, \citenamefont {Feng}, \citenamefont {Xu},\ and\
  \citenamefont {Yao}}]{Xiao:2012a}%
  \BibitemOpen
  \bibfield  {author} {\bibinfo {author} {\bibfnamefont {D.}~\bibnamefont
  {Xiao}}, \bibinfo {author} {\bibfnamefont {G.-B.}\ \bibnamefont {Liu}},
  \bibinfo {author} {\bibfnamefont {W.}~\bibnamefont {Feng}}, \bibinfo {author}
  {\bibfnamefont {X.}~\bibnamefont {Xu}},\ and\ \bibinfo {author}
  {\bibfnamefont {W.}~\bibnamefont {Yao}},\ }\bibfield  {title} {\bibinfo
  {title} {Coupled spin and valley physics in monolayers of {M}o{S}2 and other
  group-vi dichalcogenides},\ }\href@noop {} {\bibfield  {journal} {\bibinfo
  {journal} {Phys. Rev. Lett.}\ }\textbf {\bibinfo {volume} {108}},\ \bibinfo
  {pages} {196802} (\bibinfo {year} {2012})}\BibitemShut {NoStop}%
\bibitem [{\citenamefont {Li}\ \emph {et~al.}(2020{\natexlab{a}})\citenamefont
  {Li}, \citenamefont {Shao}, \citenamefont {Liu}, \citenamefont {Gao},
  \citenamefont {Wang}, \citenamefont {Zheng}, \citenamefont {Hou},
  \citenamefont {Shehzad}, \citenamefont {Yu}, \citenamefont {Miao},
  \citenamefont {Shi}, \citenamefont {Xu},\ and\ \citenamefont
  {Wang}}]{Li:2020a}%
  \BibitemOpen
  \bibfield  {author} {\bibinfo {author} {\bibfnamefont {L.}~\bibnamefont
  {Li}}, \bibinfo {author} {\bibfnamefont {L.}~\bibnamefont {Shao}}, \bibinfo
  {author} {\bibfnamefont {X.}~\bibnamefont {Liu}}, \bibinfo {author}
  {\bibfnamefont {A.}~\bibnamefont {Gao}}, \bibinfo {author} {\bibfnamefont
  {H.}~\bibnamefont {Wang}}, \bibinfo {author} {\bibfnamefont {B.}~\bibnamefont
  {Zheng}}, \bibinfo {author} {\bibfnamefont {G.}~\bibnamefont {Hou}}, \bibinfo
  {author} {\bibfnamefont {K.}~\bibnamefont {Shehzad}}, \bibinfo {author}
  {\bibfnamefont {L.}~\bibnamefont {Yu}}, \bibinfo {author} {\bibfnamefont
  {F.}~\bibnamefont {Miao}}, \bibinfo {author} {\bibfnamefont {Y.}~\bibnamefont
  {Shi}}, \bibinfo {author} {\bibfnamefont {Y.}~\bibnamefont {Xu}},\ and\
  \bibinfo {author} {\bibfnamefont {X.}~\bibnamefont {Wang}},\ }\bibfield
  {title} {\bibinfo {title} {Room-temperature valleytronic transistor,},\
  }\href@noop {} {\bibfield  {journal} {\bibinfo  {journal} {Nature
  nanotechnology}\ }\textbf {\bibinfo {volume} {15}},\ \bibinfo {pages} {743}
  (\bibinfo {year} {2020}{\natexlab{a}})}\BibitemShut {NoStop}%
\bibitem [{\citenamefont {Li}\ \emph {et~al.}(2020{\natexlab{b}})\citenamefont
  {Li}, \citenamefont {Jiang}, \citenamefont {Wang}, \citenamefont {Watanabe},
  \citenamefont {Taniguchi}, \citenamefont {Shan},\ and\ \citenamefont
  {Mak}}]{Li:2020b}%
  \BibitemOpen
  \bibfield  {author} {\bibinfo {author} {\bibfnamefont {L.}~\bibnamefont
  {Li}}, \bibinfo {author} {\bibfnamefont {S.}~\bibnamefont {Jiang}}, \bibinfo
  {author} {\bibfnamefont {Z.}~\bibnamefont {Wang}}, \bibinfo {author}
  {\bibfnamefont {K.}~\bibnamefont {Watanabe}}, \bibinfo {author}
  {\bibfnamefont {T.}~\bibnamefont {Taniguchi}}, \bibinfo {author}
  {\bibfnamefont {J.}~\bibnamefont {Shan}},\ and\ \bibinfo {author}
  {\bibfnamefont {K.~F.}\ \bibnamefont {Mak}},\ }\bibfield  {title} {\bibinfo
  {title} {Electrical switching of valley polarization in monolayer
  semiconductors},\ }\href@noop {} {\bibfield  {journal} {\bibinfo  {journal}
  {Physical Review Materials}\ }\textbf {\bibinfo {volume} {4}},\ \bibinfo
  {pages} {104005} (\bibinfo {year} {2020}{\natexlab{b}})}\BibitemShut
  {NoStop}%
\bibitem [{\citenamefont {Huang}\ \emph {et~al.}(2020)\citenamefont {Huang},
  \citenamefont {Liu}, \citenamefont {Dini}, \citenamefont {Tan}, \citenamefont
  {Liu}, \citenamefont {Fang}, \citenamefont {Liu}, \citenamefont {Liew},\ and\
  \citenamefont {Gao}}]{Huang:2020a}%
  \BibitemOpen
  \bibfield  {author} {\bibinfo {author} {\bibfnamefont {Z.}~\bibnamefont
  {Huang}}, \bibinfo {author} {\bibfnamefont {Y.}~\bibnamefont {Liu}}, \bibinfo
  {author} {\bibfnamefont {K.}~\bibnamefont {Dini}}, \bibinfo {author}
  {\bibfnamefont {.}~\bibnamefont {Tan}}, \bibinfo {author} {\bibfnamefont
  {Z.}~\bibnamefont {Liu}}, \bibinfo {author} {\bibfnamefont {H.}~\bibnamefont
  {Fang}}, \bibinfo {author} {\bibfnamefont {J.}~\bibnamefont {Liu}}, \bibinfo
  {author} {\bibfnamefont {T.}~\bibnamefont {Liew}},\ and\ \bibinfo {author}
  {\bibfnamefont {W.}~\bibnamefont {Gao}},\ }\bibfield  {title} {\bibinfo
  {title} {Robust room temperature valley hall effect of interlayer excitons},\
  }\href@noop {} {\bibfield  {journal} {\bibinfo  {journal} {Nano Letters}\
  }\textbf {\bibinfo {volume} {20}},\ \bibinfo {pages} {1345} (\bibinfo {year}
  {2020})}\BibitemShut {NoStop}%
\bibitem [{\citenamefont {Unuchek}\ \emph {et~al.}(2018)\citenamefont
  {Unuchek}, \citenamefont {Ciarrocchi}, \citenamefont {Avsar}, \citenamefont
  {Watanabe}, \citenamefont {Taniguchi},\ and\ \citenamefont
  {Kis}}]{Unichek:2018a}%
  \BibitemOpen
  \bibfield  {author} {\bibinfo {author} {\bibfnamefont {D.}~\bibnamefont
  {Unuchek}}, \bibinfo {author} {\bibfnamefont {A.}~\bibnamefont {Ciarrocchi}},
  \bibinfo {author} {\bibfnamefont {A.}~\bibnamefont {Avsar}}, \bibinfo
  {author} {\bibfnamefont {K.}~\bibnamefont {Watanabe}}, \bibinfo {author}
  {\bibfnamefont {T.}~\bibnamefont {Taniguchi}},\ and\ \bibinfo {author}
  {\bibfnamefont {A.}~\bibnamefont {Kis}},\ }\bibfield  {title} {\bibinfo
  {title} {Room-temperature electrical control of exciton flux in a van der
  waals heterostructure},\ }\href@noop {} {\bibfield  {journal} {\bibinfo
  {journal} {Nature}\ }\textbf {\bibinfo {volume} {560}},\ \bibinfo {pages}
  {340} (\bibinfo {year} {2018})}\BibitemShut {NoStop}%
\bibitem [{\citenamefont {Kulig}\ \emph {et~al.}(2018)\citenamefont {Kulig},
  \citenamefont {Zipfel}, \citenamefont {Nagler}, \citenamefont {Blanter},
  \citenamefont {Schuller}, \citenamefont {Korn}, \citenamefont {Paradiso},
  \citenamefont {Glazov},\ and\ \citenamefont {Chernikov}}]{Kulig:2018a}%
  \BibitemOpen
  \bibfield  {author} {\bibinfo {author} {\bibfnamefont {M.}~\bibnamefont
  {Kulig}}, \bibinfo {author} {\bibfnamefont {J.}~\bibnamefont {Zipfel}},
  \bibinfo {author} {\bibfnamefont {P.}~\bibnamefont {Nagler}}, \bibinfo
  {author} {\bibfnamefont {S.}~\bibnamefont {Blanter}}, \bibinfo {author}
  {\bibfnamefont {C.}~\bibnamefont {Schuller}}, \bibinfo {author}
  {\bibfnamefont {T.}~\bibnamefont {Korn}}, \bibinfo {author} {\bibfnamefont
  {N.}~\bibnamefont {Paradiso}}, \bibinfo {author} {\bibfnamefont {M.~M.}\
  \bibnamefont {Glazov}},\ and\ \bibinfo {author} {\bibfnamefont
  {A.}~\bibnamefont {Chernikov}},\ }\bibfield  {title} {\bibinfo {title}
  {Exciton diffusion and halo effects in monolayer semiconductors},\
  }\href@noop {} {\bibfield  {journal} {\bibinfo  {journal} {Physical Review
  Letters}\ }\textbf {\bibinfo {volume} {120}},\ \bibinfo {pages} {207401}
  (\bibinfo {year} {2018})}\BibitemShut {NoStop}%
\bibitem [{\citenamefont {Perea-Causin}\ \emph {et~al.}(2019)\citenamefont
  {Perea-Causin}, \citenamefont {Brem}, \citenamefont {Rosati}, \citenamefont
  {Jago}, \citenamefont {Kulig}, \citenamefont {Ziegler}, \citenamefont
  {Zipfel}, \citenamefont {Chernikov},\ and\ \citenamefont
  {Malic}}]{Perea:2019a}%
  \BibitemOpen
  \bibfield  {author} {\bibinfo {author} {\bibfnamefont {R.}~\bibnamefont
  {Perea-Causin}}, \bibinfo {author} {\bibfnamefont {S.}~\bibnamefont {Brem}},
  \bibinfo {author} {\bibfnamefont {R.}~\bibnamefont {Rosati}}, \bibinfo
  {author} {\bibfnamefont {R.}~\bibnamefont {Jago}}, \bibinfo {author}
  {\bibfnamefont {M.}~\bibnamefont {Kulig}}, \bibinfo {author} {\bibfnamefont
  {J.}~\bibnamefont {Ziegler}}, \bibinfo {author} {\bibfnamefont
  {J.}~\bibnamefont {Zipfel}}, \bibinfo {author} {\bibfnamefont
  {A.}~\bibnamefont {Chernikov}},\ and\ \bibinfo {author} {\bibfnamefont
  {E.}~\bibnamefont {Malic}},\ }\bibfield  {title} {\bibinfo {title} {Exciton
  propagation and halo formation in two-dimensional materials},\ }\href@noop {}
  {\bibfield  {journal} {\bibinfo  {journal} {Nano Letters}\ }\textbf {\bibinfo
  {volume} {19}},\ \bibinfo {pages} {7317} (\bibinfo {year}
  {2019})}\BibitemShut {NoStop}%
\bibitem [{\citenamefont {{Cadiz}}\ \emph {et~al.}(2017)\citenamefont
  {{Cadiz}}, \citenamefont {{Courtade}}, \citenamefont {{Robert}},
  \citenamefont {{Wang}}, \citenamefont {{Shen}}, \citenamefont {{Cai}},
  \citenamefont {{Taniguchi}}, \citenamefont {{Watanabe}}, \citenamefont
  {{Carrere}}, \citenamefont {{Lagarde}}, \citenamefont {{Manca}},
  \citenamefont {{Amand}}, \citenamefont {{Renucci}}, \citenamefont {{Tongay}},
  \citenamefont {{Marie}},\ and\ \citenamefont {{Urbaszek}}}]{Cadiz:2017a}%
  \BibitemOpen
  \bibfield  {author} {\bibinfo {author} {\bibfnamefont {F.}~\bibnamefont
  {{Cadiz}}}, \bibinfo {author} {\bibfnamefont {E.}~\bibnamefont {{Courtade}}},
  \bibinfo {author} {\bibfnamefont {C.}~\bibnamefont {{Robert}}}, \bibinfo
  {author} {\bibfnamefont {G.}~\bibnamefont {{Wang}}}, \bibinfo {author}
  {\bibfnamefont {Y.}~\bibnamefont {{Shen}}}, \bibinfo {author} {\bibfnamefont
  {H.}~\bibnamefont {{Cai}}}, \bibinfo {author} {\bibfnamefont
  {T.}~\bibnamefont {{Taniguchi}}}, \bibinfo {author} {\bibfnamefont
  {K.}~\bibnamefont {{Watanabe}}}, \bibinfo {author} {\bibfnamefont
  {H.}~\bibnamefont {{Carrere}}}, \bibinfo {author} {\bibfnamefont
  {D.}~\bibnamefont {{Lagarde}}}, \bibinfo {author} {\bibfnamefont
  {M.}~\bibnamefont {{Manca}}}, \bibinfo {author} {\bibfnamefont
  {T.}~\bibnamefont {{Amand}}}, \bibinfo {author} {\bibfnamefont
  {P.}~\bibnamefont {{Renucci}}}, \bibinfo {author} {\bibfnamefont
  {S.}~\bibnamefont {{Tongay}}}, \bibinfo {author} {\bibfnamefont
  {X.}~\bibnamefont {{Marie}}},\ and\ \bibinfo {author} {\bibfnamefont
  {B.}~\bibnamefont {{Urbaszek}}},\ }\bibfield  {title} {\bibinfo {title}
  {{Excitonic linewidth approaching the homogeneous limit in MoS2 based van der
  Waals heterostructures : accessing spin-valley dynamics}},\ }\href@noop {}
  {\bibfield  {journal} {\bibinfo  {journal} {Phys. Rev. X}\ }\textbf {\bibinfo
  {volume} {7}},\ \bibinfo {pages} {021026} (\bibinfo {year} {2017})},\ \Eprint
  {https://arxiv.org/abs/1702.00323} {1702.00323} \BibitemShut {NoStop}%
\bibitem [{\citenamefont {Castellanos-Gomez}\ \emph {et~al.}(2014)\citenamefont
  {Castellanos-Gomez}, \citenamefont {Buscema}, \citenamefont {Molenaar},
  \citenamefont {Singh}, \citenamefont {Janssen}, \citenamefont {van~der
  Zant},\ and\ \citenamefont {Steele}}]{Gomez:2014a}%
  \BibitemOpen
  \bibfield  {author} {\bibinfo {author} {\bibfnamefont {A.}~\bibnamefont
  {Castellanos-Gomez}}, \bibinfo {author} {\bibfnamefont {M.}~\bibnamefont
  {Buscema}}, \bibinfo {author} {\bibfnamefont {R.}~\bibnamefont {Molenaar}},
  \bibinfo {author} {\bibfnamefont {V.}~\bibnamefont {Singh}}, \bibinfo
  {author} {\bibfnamefont {L.}~\bibnamefont {Janssen}}, \bibinfo {author}
  {\bibfnamefont {H.~S.~J.}\ \bibnamefont {van~der Zant}},\ and\ \bibinfo
  {author} {\bibfnamefont {G.~A.}\ \bibnamefont {Steele}},\ }\bibfield  {title}
  {\bibinfo {title} {Deterministic transfer of two-dimensional materials by
  all-dry viscoelastic stamping},\ }\href@noop {} {\bibfield  {journal}
  {\bibinfo  {journal} {2D Materials}\ }\textbf {\bibinfo {volume} {1}},\
  \bibinfo {pages} {011002} (\bibinfo {year} {2014})}\BibitemShut {NoStop}%
\bibitem [{\citenamefont {Favorskiy}\ \emph {et~al.}(2010)\citenamefont
  {Favorskiy}, \citenamefont {Vu}, \citenamefont {Peytavit}, \citenamefont
  {Arscott}, \citenamefont {Paget},\ and\ \citenamefont
  {Rowe}}]{Favorskiy:2010a}%
  \BibitemOpen
  \bibfield  {author} {\bibinfo {author} {\bibfnamefont {I.}~\bibnamefont
  {Favorskiy}}, \bibinfo {author} {\bibfnamefont {D.}~\bibnamefont {Vu}},
  \bibinfo {author} {\bibfnamefont {E.}~\bibnamefont {Peytavit}}, \bibinfo
  {author} {\bibfnamefont {S.}~\bibnamefont {Arscott}}, \bibinfo {author}
  {\bibfnamefont {D.}~\bibnamefont {Paget}},\ and\ \bibinfo {author}
  {\bibfnamefont {A.~C.~H.}\ \bibnamefont {Rowe}},\ }\bibfield  {title}
  {\bibinfo {title} {Circularly polarized luminescence mmicroscopy for the
  imaging of charge and spin diffusion in semiconductors},\ }\href@noop {}
  {\bibfield  {journal} {\bibinfo  {journal} {Review of Scientific
  Instruments}\ }\textbf {\bibinfo {volume} {81}},\ \bibinfo {pages} {103902}
  (\bibinfo {year} {2010})}\BibitemShut {NoStop}%
\bibitem [{\citenamefont {Cadiz}\ \emph {et~al.}(2018)\citenamefont {Cadiz},
  \citenamefont {Robert}, \citenamefont {Courtade}, \citenamefont {Manca},
  \citenamefont {Martinelli}, \citenamefont {Taniguchi}, \citenamefont
  {Watanabe}, \citenamefont {Amand}, \citenamefont {Rowe}, \citenamefont
  {Paget}, \citenamefont {Urbaszek},\ and\ \citenamefont
  {Marie}}]{Cadiz:2018a}%
  \BibitemOpen
  \bibfield  {author} {\bibinfo {author} {\bibfnamefont {F.}~\bibnamefont
  {Cadiz}}, \bibinfo {author} {\bibfnamefont {C.}~\bibnamefont {Robert}},
  \bibinfo {author} {\bibfnamefont {E.}~\bibnamefont {Courtade}}, \bibinfo
  {author} {\bibfnamefont {M.}~\bibnamefont {Manca}}, \bibinfo {author}
  {\bibfnamefont {L.}~\bibnamefont {Martinelli}}, \bibinfo {author}
  {\bibfnamefont {T.}~\bibnamefont {Taniguchi}}, \bibinfo {author}
  {\bibfnamefont {K.}~\bibnamefont {Watanabe}}, \bibinfo {author}
  {\bibfnamefont {T.}~\bibnamefont {Amand}}, \bibinfo {author} {\bibfnamefont
  {A.}~\bibnamefont {Rowe}}, \bibinfo {author} {\bibfnamefont {D.}~\bibnamefont
  {Paget}}, \bibinfo {author} {\bibfnamefont {B.}~\bibnamefont {Urbaszek}},\
  and\ \bibinfo {author} {\bibfnamefont {X.}~\bibnamefont {Marie}},\ }\bibfield
   {title} {\bibinfo {title} {Exciton diffusion in {W}{S}e2 monolayers embedded
  in a van der waals heterostructure},\ }\href@noop {} {\bibfield  {journal}
  {\bibinfo  {journal} {Applied Physics Letters}\ } (\bibinfo {year}
  {2018})}\BibitemShut {NoStop}%
\bibitem [{\citenamefont {Cadiz}\ \emph {et~al.}(2016)\citenamefont {Cadiz},
  \citenamefont {Robert}, \citenamefont {Wang}, \citenamefont {Kong},
  \citenamefont {Fan}, \citenamefont {Blei}, \citenamefont {Lagarde},
  \citenamefont {Gay}, \citenamefont {Manca}, \citenamefont {Taniguchi},
  \citenamefont {Watanabe}, \citenamefont {Amand}, \citenamefont {Marie},
  \citenamefont {Renucci}, \citenamefont {Tongay},\ and\ \citenamefont
  {Urbaszek}}]{Cadiz:2016b}%
  \BibitemOpen
  \bibfield  {author} {\bibinfo {author} {\bibfnamefont {F.}~\bibnamefont
  {Cadiz}}, \bibinfo {author} {\bibfnamefont {C.}~\bibnamefont {Robert}},
  \bibinfo {author} {\bibfnamefont {G.}~\bibnamefont {Wang}}, \bibinfo {author}
  {\bibfnamefont {W.}~\bibnamefont {Kong}}, \bibinfo {author} {\bibfnamefont
  {X.}~\bibnamefont {Fan}}, \bibinfo {author} {\bibfnamefont {M.}~\bibnamefont
  {Blei}}, \bibinfo {author} {\bibfnamefont {D.}~\bibnamefont {Lagarde}},
  \bibinfo {author} {\bibfnamefont {M.}~\bibnamefont {Gay}}, \bibinfo {author}
  {\bibfnamefont {M.}~\bibnamefont {Manca}}, \bibinfo {author} {\bibfnamefont
  {T.}~\bibnamefont {Taniguchi}}, \bibinfo {author} {\bibfnamefont
  {K.}~\bibnamefont {Watanabe}}, \bibinfo {author} {\bibfnamefont
  {T.}~\bibnamefont {Amand}}, \bibinfo {author} {\bibfnamefont
  {X.}~\bibnamefont {Marie}}, \bibinfo {author} {\bibfnamefont
  {P.}~\bibnamefont {Renucci}}, \bibinfo {author} {\bibfnamefont
  {S.}~\bibnamefont {Tongay}},\ and\ \bibinfo {author} {\bibfnamefont
  {B.}~\bibnamefont {Urbaszek}},\ }\bibfield  {title} {\bibinfo {title}
  {Ultra-low power threshold for laser induced changes in optical properties of
  2{D} molybdenum dichalcogenides},\ }\href@noop {} {\bibfield  {journal}
  {\bibinfo  {journal} {2D Materials}\ }\textbf {\bibinfo {volume} {3}},\
  \bibinfo {pages} {045008} (\bibinfo {year} {2016})}\BibitemShut {NoStop}%
\bibitem [{\citenamefont {Fang}\ \emph {et~al.}(2019)\citenamefont {Fang},
  \citenamefont {Han}, \citenamefont {Robert}, \citenamefont {Semina},
  \citenamefont {Lagarde}, \citenamefont {Courtade}, \citenamefont {Taniguchi},
  \citenamefont {Watanabe}, \citenamefont {Amand}, \citenamefont {Urbaszek},
  \citenamefont {Glazov},\ and\ \citenamefont {Marie}}]{Fang:2019a}%
  \BibitemOpen
  \bibfield  {author} {\bibinfo {author} {\bibfnamefont {H.~H.}\ \bibnamefont
  {Fang}}, \bibinfo {author} {\bibfnamefont {B.}~\bibnamefont {Han}}, \bibinfo
  {author} {\bibfnamefont {C.}~\bibnamefont {Robert}}, \bibinfo {author}
  {\bibfnamefont {M.~A.}\ \bibnamefont {Semina}}, \bibinfo {author}
  {\bibfnamefont {D.}~\bibnamefont {Lagarde}}, \bibinfo {author} {\bibfnamefont
  {E.}~\bibnamefont {Courtade}}, \bibinfo {author} {\bibfnamefont
  {T.}~\bibnamefont {Taniguchi}}, \bibinfo {author} {\bibfnamefont
  {K.}~\bibnamefont {Watanabe}}, \bibinfo {author} {\bibfnamefont
  {T.}~\bibnamefont {Amand}}, \bibinfo {author} {\bibfnamefont
  {B.}~\bibnamefont {Urbaszek}}, \bibinfo {author} {\bibfnamefont {M.~M.}\
  \bibnamefont {Glazov}},\ and\ \bibinfo {author} {\bibfnamefont
  {X.}~\bibnamefont {Marie}},\ }\bibfield  {title} {\bibinfo {title} {Control
  of the exciton radiative lifetime in van der waals heterostructures},\
  }\href@noop {} {\bibfield  {journal} {\bibinfo  {journal} {Physical Review
  Letters}\ }\textbf {\bibinfo {volume} {123}},\ \bibinfo {pages} {067401}
  (\bibinfo {year} {2019})}\BibitemShut {NoStop}%
\bibitem [{\citenamefont {Hoshi}\ \emph {et~al.}(2017)\citenamefont {Hoshi},
  \citenamefont {Kuroda}, \citenamefont {Okada}, \citenamefont {Moriya},
  \citenamefont {Masubichi}, \citenamefont {Watanabe}, \citenamefont
  {Taniguchi}, \citenamefont {Kitaura},\ and\ \citenamefont
  {Machida}}]{Hoshi:2017a}%
  \BibitemOpen
  \bibfield  {author} {\bibinfo {author} {\bibfnamefont {Y.}~\bibnamefont
  {Hoshi}}, \bibinfo {author} {\bibfnamefont {T.}~\bibnamefont {Kuroda}},
  \bibinfo {author} {\bibfnamefont {M.}~\bibnamefont {Okada}}, \bibinfo
  {author} {\bibfnamefont {R.}~\bibnamefont {Moriya}}, \bibinfo {author}
  {\bibfnamefont {S.}~\bibnamefont {Masubichi}}, \bibinfo {author}
  {\bibfnamefont {K.}~\bibnamefont {Watanabe}}, \bibinfo {author}
  {\bibfnamefont {T.}~\bibnamefont {Taniguchi}}, \bibinfo {author}
  {\bibfnamefont {R.}~\bibnamefont {Kitaura}},\ and\ \bibinfo {author}
  {\bibfnamefont {T.}~\bibnamefont {Machida}},\ }\bibfield  {title} {\bibinfo
  {title} {Suppression of exciton-exciton annihilation in tungsten disulfide
  monolayers encapsulated by hexagonal boron nitrides},\ }\href@noop {}
  {\bibfield  {journal} {\bibinfo  {journal} {Phys. Rev. B}\ }\textbf {\bibinfo
  {volume} {95}},\ \bibinfo {pages} {241403(R)} (\bibinfo {year}
  {2017})}\BibitemShut {NoStop}%
\bibitem [{\citenamefont {Zipfel}\ \emph {et~al.}(2020)\citenamefont {Zipfel},
  \citenamefont {Kulig}, \citenamefont {Perea-Causin}, \citenamefont {Brem},
  \citenamefont {Ziegler}, \citenamefont {Rosati}, \citenamefont {Taniguchi},
  \citenamefont {Watanabe}, \citenamefont {Glazov}, \citenamefont {Malic},\
  and\ \citenamefont {Chernikov}}]{Zipfel:2020a}%
  \BibitemOpen
  \bibfield  {author} {\bibinfo {author} {\bibfnamefont {J.}~\bibnamefont
  {Zipfel}}, \bibinfo {author} {\bibfnamefont {M.}~\bibnamefont {Kulig}},
  \bibinfo {author} {\bibfnamefont {R.}~\bibnamefont {Perea-Causin}}, \bibinfo
  {author} {\bibfnamefont {S.}~\bibnamefont {Brem}}, \bibinfo {author}
  {\bibfnamefont {J.}~\bibnamefont {Ziegler}}, \bibinfo {author} {\bibfnamefont
  {R.}~\bibnamefont {Rosati}}, \bibinfo {author} {\bibfnamefont
  {T.}~\bibnamefont {Taniguchi}}, \bibinfo {author} {\bibfnamefont
  {K.}~\bibnamefont {Watanabe}}, \bibinfo {author} {\bibfnamefont {M.~M.}\
  \bibnamefont {Glazov}}, \bibinfo {author} {\bibfnamefont {E.}~\bibnamefont
  {Malic}},\ and\ \bibinfo {author} {\bibfnamefont {A.}~\bibnamefont
  {Chernikov}},\ }\bibfield  {title} {\bibinfo {title} {Exciton diffusion in
  monolayer semiconductors with suppressed disorder},\ }\href@noop {}
  {\bibfield  {journal} {\bibinfo  {journal} {Physical Review B}\ }\textbf
  {\bibinfo {volume} {101}},\ \bibinfo {pages} {115430} (\bibinfo {year}
  {2020})}\BibitemShut {NoStop}%
\bibitem [{\citenamefont {Cordovilla~Leon}\ \emph {et~al.}(2019)\citenamefont
  {Cordovilla~Leon}, \citenamefont {Li}, \citenamefont {Jang},\ and\
  \citenamefont {Deotare}}]{Cordovilla:2019a}%
  \BibitemOpen
  \bibfield  {author} {\bibinfo {author} {\bibfnamefont {D.~F.}\ \bibnamefont
  {Cordovilla~Leon}}, \bibinfo {author} {\bibfnamefont {Z.}~\bibnamefont {Li}},
  \bibinfo {author} {\bibfnamefont {S.~W.}\ \bibnamefont {Jang}},\ and\
  \bibinfo {author} {\bibfnamefont {P.~B.}\ \bibnamefont {Deotare}},\
  }\bibfield  {title} {\bibinfo {title} {Hot exciton transpot in {W}{S}e2
  monolayers},\ }\href@noop {} {\bibfield  {journal} {\bibinfo  {journal}
  {Physical Review B}\ }\textbf {\bibinfo {volume} {100}},\ \bibinfo {pages}
  {241401(R)} (\bibinfo {year} {2019})}\BibitemShut {NoStop}%
\bibitem [{\citenamefont {Christopher}\ \emph {et~al.}(2017)\citenamefont
  {Christopher}, \citenamefont {Goldberg},\ and\ \citenamefont
  {Swan}}]{Christopher:2017a}%
  \BibitemOpen
  \bibfield  {author} {\bibinfo {author} {\bibfnamefont {J.~W.}\ \bibnamefont
  {Christopher}}, \bibinfo {author} {\bibfnamefont {B.~B.}\ \bibnamefont
  {Goldberg}},\ and\ \bibinfo {author} {\bibfnamefont {A.~K.}\ \bibnamefont
  {Swan}},\ }\bibfield  {title} {\bibinfo {title} {Long tailed trions in
  monolayer {M}o{S}2: Temperature dependent asymmetry and resulting red-shift
  of trion photoluminescence spectra},\ }\href@noop {} {\bibfield  {journal}
  {\bibinfo  {journal} {Scientific Reports}\ }\textbf {\bibinfo {volume} {7}},\
  \bibinfo {pages} {14062} (\bibinfo {year} {2017})}\BibitemShut {NoStop}%
\bibitem [{\citenamefont {Zhumagulov}\ \emph {et~al.}(2020)\citenamefont
  {Zhumagulov}, \citenamefont {Vagov}, \citenamefont {Gulevich}, \citenamefont
  {Faria~Junio},\ and\ \citenamefont {Perebeinos}}]{Zhumagulov:2020a}%
  \BibitemOpen
  \bibfield  {author} {\bibinfo {author} {\bibfnamefont {Y.}~\bibnamefont
  {Zhumagulov}}, \bibinfo {author} {\bibfnamefont {A.}~\bibnamefont {Vagov}},
  \bibinfo {author} {\bibfnamefont {D.~R.}\ \bibnamefont {Gulevich}}, \bibinfo
  {author} {\bibfnamefont {P.~E.}\ \bibnamefont {Faria~Junio}},\ and\ \bibinfo
  {author} {\bibfnamefont {V.}~\bibnamefont {Perebeinos}},\ }\bibfield  {title}
  {\bibinfo {title} {Trion induced photoluminescence of a doped {M}o{S}2
  monolayer},\ }\href@noop {} {\bibfield  {journal} {\bibinfo  {journal} {J.
  Chem. Phys.}\ }\textbf {\bibinfo {volume} {153}},\ \bibinfo {pages} {044132}
  (\bibinfo {year} {2020})}\BibitemShut {NoStop}%
\bibitem [{\citenamefont {Madeo}\ \emph {et~al.}(2020)\citenamefont {Madeo},
  \citenamefont {Man}, \citenamefont {Sahoo}, \citenamefont {Campbell},
  \citenamefont {Pareek}, \citenamefont {Wong}, \citenamefont {Al-Mahboob},
  \citenamefont {Chan}, \citenamefont {Karmakar}, \citenamefont
  {Krishna~Mariserla}, \citenamefont {Li}, \citenamefont {Heinz}, \citenamefont
  {Cao},\ and\ \citenamefont {Dani}}]{Madeo:2020a}%
  \BibitemOpen
  \bibfield  {author} {\bibinfo {author} {\bibfnamefont {J.}~\bibnamefont
  {Madeo}}, \bibinfo {author} {\bibfnamefont {M.}~\bibnamefont {Man}}, \bibinfo
  {author} {\bibfnamefont {C.}~\bibnamefont {Sahoo}}, \bibinfo {author}
  {\bibfnamefont {M.}~\bibnamefont {Campbell}}, \bibinfo {author}
  {\bibfnamefont {V.}~\bibnamefont {Pareek}}, \bibinfo {author} {\bibfnamefont
  {E.}~\bibnamefont {Wong}}, \bibinfo {author} {\bibfnamefont {A.}~\bibnamefont
  {Al-Mahboob}}, \bibinfo {author} {\bibfnamefont {N.}~\bibnamefont {Chan}},
  \bibinfo {author} {\bibfnamefont {A.}~\bibnamefont {Karmakar}}, \bibinfo
  {author} {\bibfnamefont {B.~M.}\ \bibnamefont {Krishna~Mariserla}}, \bibinfo
  {author} {\bibfnamefont {X.}~\bibnamefont {Li}}, \bibinfo {author}
  {\bibfnamefont {T.~F.}\ \bibnamefont {Heinz}}, \bibinfo {author}
  {\bibfnamefont {T.}~\bibnamefont {Cao}},\ and\ \bibinfo {author}
  {\bibfnamefont {K.~M.}\ \bibnamefont {Dani}},\ }\bibfield  {title} {\bibinfo
  {title} {Directly visualizing the momentum-fordbidden dark excitons and their
  dynamics in atomically thin semiconductors},\ }\href@noop {} {\bibfield
  {journal} {\bibinfo  {journal} {Science}\ }\textbf {\bibinfo {volume}
  {370}},\ \bibinfo {pages} {1199} (\bibinfo {year} {2020})}\BibitemShut
  {NoStop}%
\bibitem [{\citenamefont {Cai}\ and\ \citenamefont {Mahan}()}]{Cai:2006a}%
  \BibitemOpen
  \bibfield  {author} {\bibinfo {author} {\bibfnamefont {J.}~\bibnamefont
  {Cai}}\ and\ \bibinfo {author} {\bibfnamefont {G.~D.}\ \bibnamefont
  {Mahan}},\ }\bibfield  {title} {\bibinfo {title} {Effective seebeck
  coefficient for semiconductors},\ }\href@noop {} {\bibfield  {journal}
  {\bibinfo  {journal} {Physical Review B}\ }\textbf {\bibinfo {volume} {74}},\
  \bibinfo {pages} {075201}}\BibitemShut {NoStop}%
\bibitem [{\citenamefont {Bieker}\ \emph {et~al.}(2015)\citenamefont {Bieker},
  \citenamefont {Henn}, \citenamefont {Kiessling}, \citenamefont {Ossau},\ and\
  \citenamefont {Molenkamp}}]{Bieker:2015a}%
  \BibitemOpen
  \bibfield  {author} {\bibinfo {author} {\bibfnamefont {S.}~\bibnamefont
  {Bieker}}, \bibinfo {author} {\bibfnamefont {T.}~\bibnamefont {Henn}},
  \bibinfo {author} {\bibfnamefont {T.}~\bibnamefont {Kiessling}}, \bibinfo
  {author} {\bibfnamefont {W.}~\bibnamefont {Ossau}},\ and\ \bibinfo {author}
  {\bibfnamefont {L.~W.}\ \bibnamefont {Molenkamp}},\ }\bibfield  {title}
  {\bibinfo {title} {Spatially resolved thermodynamics of the partially ionized
  exciton gas in {G}a{A}s},\ }\href@noop {} {\bibfield  {journal} {\bibinfo
  {journal} {Physical Review Letters}\ }\textbf {\bibinfo {volume} {114}},\
  \bibinfo {pages} {227402} (\bibinfo {year} {2015})}\BibitemShut {NoStop}%
\bibitem [{\citenamefont {Han}\ \emph {et~al.}()\citenamefont {Han},
  \citenamefont {Robert}, \citenamefont {Courtade}, \citenamefont {Manca},
  \citenamefont {Shree}, \citenamefont {Amand}, \citenamefont {Renucci},
  \citenamefont {Taniguchi}, \citenamefont {Watanabe}, \citenamefont {Marie},
  \citenamefont {Golub}, \citenamefont {Glazov},\ and\ \citenamefont
  {Urbaszek}}]{Han:2018a}%
  \BibitemOpen
  \bibfield  {author} {\bibinfo {author} {\bibfnamefont {B.}~\bibnamefont
  {Han}}, \bibinfo {author} {\bibfnamefont {C.}~\bibnamefont {Robert}},
  \bibinfo {author} {\bibfnamefont {E.}~\bibnamefont {Courtade}}, \bibinfo
  {author} {\bibfnamefont {M.}~\bibnamefont {Manca}}, \bibinfo {author}
  {\bibfnamefont {S.}~\bibnamefont {Shree}}, \bibinfo {author} {\bibfnamefont
  {T.}~\bibnamefont {Amand}}, \bibinfo {author} {\bibfnamefont
  {P.}~\bibnamefont {Renucci}}, \bibinfo {author} {\bibfnamefont
  {T.}~\bibnamefont {Taniguchi}}, \bibinfo {author} {\bibfnamefont
  {K.}~\bibnamefont {Watanabe}}, \bibinfo {author} {\bibfnamefont
  {X.}~\bibnamefont {Marie}}, \bibinfo {author} {\bibfnamefont {L.~E.}\
  \bibnamefont {Golub}}, \bibinfo {author} {\bibfnamefont {M.~M.}\ \bibnamefont
  {Glazov}},\ and\ \bibinfo {author} {\bibfnamefont {B.}~\bibnamefont
  {Urbaszek}},\ }\bibfield  {title} {\bibinfo {title} {Exciton states in
  monolayer {M}o{S}e2 and {M}o{T}e2 probed by upconversion spectroscopy},\
  }\href@noop {} {\bibfield  {journal} {\bibinfo  {journal} {Phys. Rev. X}\
  }\textbf {\bibinfo {volume} {8}},\ \bibinfo {pages} {031073}}\BibitemShut
  {NoStop}%
\bibitem [{\citenamefont {Sie}\ \emph {et~al.}(2017)\citenamefont {Sie},
  \citenamefont {Steinhoff}, \citenamefont {Gies}, \citenamefont {Lui},
  \citenamefont {Ma}, \citenamefont {Rosner}, \citenamefont {Schonhoff},
  \citenamefont {Jhanke}, \citenamefont {Wehling}, \citenamefont {Lee},
  \citenamefont {Jong}, \citenamefont {Jarillo-Herrero},\ and\ \citenamefont
  {Gedik}}]{Sie:2017a}%
  \BibitemOpen
  \bibfield  {author} {\bibinfo {author} {\bibfnamefont {E.~J.}\ \bibnamefont
  {Sie}}, \bibinfo {author} {\bibfnamefont {A.}~\bibnamefont {Steinhoff}},
  \bibinfo {author} {\bibfnamefont {C.}~\bibnamefont {Gies}}, \bibinfo {author}
  {\bibfnamefont {C.}~\bibnamefont {Lui}}, \bibinfo {author} {\bibfnamefont
  {Q.}~\bibnamefont {Ma}}, \bibinfo {author} {\bibfnamefont {M.}~\bibnamefont
  {Rosner}}, \bibinfo {author} {\bibfnamefont {G.}~\bibnamefont {Schonhoff}},
  \bibinfo {author} {\bibfnamefont {F.}~\bibnamefont {Jhanke}}, \bibinfo
  {author} {\bibfnamefont {T.~O.}\ \bibnamefont {Wehling}}, \bibinfo {author}
  {\bibfnamefont {Y.~H.}\ \bibnamefont {Lee}}, \bibinfo {author} {\bibfnamefont
  {J.}~\bibnamefont {Jong}}, \bibinfo {author} {\bibfnamefont {P.}~\bibnamefont
  {Jarillo-Herrero}},\ and\ \bibinfo {author} {\bibfnamefont {N.}~\bibnamefont
  {Gedik}},\ }\bibfield  {title} {\bibinfo {title} {Observation of exciton
  redshift-blueshift crossover in monolayer {W}{S}2},\ }\href@noop {}
  {\bibfield  {journal} {\bibinfo  {journal} {Nano Letters}\ }\textbf {\bibinfo
  {volume} {17}},\ \bibinfo {pages} {4210} (\bibinfo {year}
  {2017})}\BibitemShut {NoStop}%
\end{thebibliography}%

\end{document}